\documentclass[10pt, conference, compsocconf]{IEEEtran}

\ifCLASSINFOpdf
\else
\fi
\usepackage{cite}
\usepackage{amsmath,amssymb,amsfonts}
\usepackage{algorithmic}
\usepackage{graphicx}
\graphicspath{{./figures/}}
\usepackage{textcomp}
\usepackage[table]{xcolor}
\usepackage{xcolor}
\usepackage{url}
\usepackage{listings}
\usepackage{colortbl}
\usepackage{array, makecell}
\usepackage{xspace}
\usepackage{caption}
\usepackage{subcaption}
\setcellgapes{2pt}
\usepackage{hyperref}
\let\labelindent\relax
\usepackage{enumitem}
\hypersetup{
    colorlinks=true,
    linkcolor=blue,
    filecolor=magenta,      
    urlcolor=cyan,
}

\newcommand{\optane}{\texttt{Optane}\xspace}

\newcommand{\registermark}{\raisebox{.5pt}{\scriptsize\textcircled{\raisebox{-.9pt} {R}}}~}
\makeatletter
\newcommand{\newlineauthors}{
\end{@IEEEauthorhalign}\hfill\mbox{}\par
\mbox{}\hfill\begin{@IEEEauthorhalign}
}
\makeatother
\def\BibTeX{{\rm B\kern-.05em{\sc i\kern-.025em b}\kern-.08em
    T\kern-.1667em\lower.7ex\hbox{E}\kern-.125emX}}

\begin{document}

\title{Demystifying the Performance of HPC Scientific Applications \\on NVM-based Memory Systems}


\author{\IEEEauthorblockN{Ivy Peng}
\IEEEauthorblockA{Lawrence Livermore National Laboratory\\
Livermore, USA\\
peng8@llnl.gov}
\and
\IEEEauthorblockN{Kai Wu}
\IEEEauthorblockA{University of California, Merced\\
Merced, USA\\ 
kwu42@ucmerced.edu}
\and
\IEEEauthorblockN{Jie Ren}
\IEEEauthorblockA{University of California, Merced\\
Merced, USA\\ 
jren6@ucmerced.edu}
\and
\newlineauthors
\IEEEauthorblockN{Dong Li}
\IEEEauthorblockA{University of California, Merced\\
Merced, USA\\ 
dli35@ucmerced.edu}
\and
\IEEEauthorblockN{Maya Gokhale}
\IEEEauthorblockA{Lawrence Livermore National Laboratory\\
Livermore, USA\\
gokhale2@llnl.gov}
}

%

\maketitle

\begin{abstract}
The emergence of high-density byte-addressable non-volatile memory (NVM) is promising to accelerate data- and compute-intensive applications. Current NVM technologies have lower performance than DRAM and, thus, are often paired with DRAM in a heterogeneous main memory. Recently, byte-addressable NVM hardware becomes available. This work provides a timely evaluation of representative HPC applications from the ``Seven Dwarfs'' on NVM-based main memory. Our results quantify the effectiveness of DRAM-cached-NVM for accelerating HPC applications and enabling large problems beyond the DRAM capacity. On uncached-NVM, HPC applications exhibit three tiers of performance sensitivity, i.e., insensitive, scaled, and bottlenecked. We identify~\emph{write throttling} and~\emph{concurrency control} as the priorities in optimizing applications. We highlight that concurrency change may have a diverging effect on read and write accesses in applications. Based on these findings, we explore two optimization approaches. First, we provide a prediction model that uses datasets from a small set of configurations to estimate performance at various concurrency and data sizes to avoid exhaustive search in the configuration space. Second, we demonstrate that write-aware data placement on uncached-NVM could achieve $2$x performance improvement with a $60\%$ reduction in DRAM usage.
\end{abstract}


\IEEEpeerreviewmaketitle

\section{Introduction}
Byte-addressable non-volatile memories, such as STT-RAM, ReRAM, and PCM~\cite{hosomi2005novel,strukov2008missing,Qureshi2009,lee2009architecting}, are promising to accelerate data- and compute-intensive HPC applications~\cite{nvm_ipdps12}. High-density NVM enables larger memory capacity than DRAM under the same area constraints. Data stored in NVM can persist through power failures as if in storage. Recently, some NVM technologies may even provide comparable bandwidth and latency to that of DRAM, enabling much higher performance than block devices. Altogether, these characteristics start blurring the boundary between memory and storage when NVM is used in the main memory. However, NVM technologies are still under active development and not ready for replacing DRAM. For instance, the write bandwidth of the Intel Optane DC persistent memory is only one third that of DRAM~\cite{peng2019}. Consequently, NVM is often paired with DRAM, building a heterogeneous memory system.  

The recent release of the Intel Optane DC persistent memory module (named \optane in the rest of the paper) marks the first mass production of byte-addressable NVM. The \optane provides a realistic and accessible hardware platform for evaluating the impact of new main memory designs on HPC scientific applications. The future Exascale system is reported to be based on an NVM technology like~\optane \cite{aurora_anl}. Therefore, the performance of HPC applications on this new hardware requires a timely and comprehensive evaluation. Several works have provided system evaluation and performance of specific applications~\cite{ucsd19,peng2019,Yang2019AnEG,utexas19}. Still, the landscape of HPC scientific applications requires a systematic approach to identify bottlenecks and opportunities. Does an NVM-based main memory change the priority in optimization? How to effectively leverage the heterogeneity in DRAM/NVM systems for the best performance? Answering these questions not only helps to exploit NVM on the next generation supercomputers but also influences the design of runtime and system software to accommodate this emerging memory technology.

In this work, we follow the well-known Seven Dwarfs~\cite{asanovic2006landscape} and choose flagship libraries and applications, such as ScaLAPACK~\cite{blackford1997scalapack}, SuperLU~\cite{li2005overview}, and Hypre~\cite{yang2006parallel} to cover the landscape of scientific applications. Our study provides a comprehensive evaluation of the domains of dense and sparse linear algebra, spectral methods, N-body methods, structured and unstructured grids, and Monte Carlo-based algorithms. We find that scientific applications exhibit three tiers of sensitivity on the uncached-NVM, i.e., insensitive, scaled, and bottlenecked. Leveraging DRAM as a cache to NVM could effectively improve application performance even when the input problems have a memory footprint three to five times the DRAM capacity. Furthermore, we reveal two bottlenecks arising from the asymmetric bandwidth and scaling limitation in NVM, i.e., \emph{write-throttling} and \emph{concurrency contention}. These characteristics may change the critical computation phases in applications and, thus, require different priorities in optimization. Also, we identify that concurrency changes have a \emph{diverging effect} on read and write accesses in applications, which requires different strategies like the write-aware placement. We believe that this work provides insights and feedbacks that are critical for applications to leverage future systems with NVM-based main memory.

We explore two optimization directions. We develop a model to predict application performance in cached-NVM at different concurrency and problem sizes to help design space exploration and identify optimal configurations. On uncached-NVM, we demonstrate in ScaLAPACK that explicitly managing write-aware placement can significantly improve performance and reduce DRAM usage. We summarize our contributions as follows.

\begin{itemize}[leftmargin=*]
\item A comprehensive performance study of HPC workloads from common computation domains (the Seven Dwarfs) on cached and uncached NVM-based main memory;
\item Highlight that write throttling and concurrency contention change the priority of optimizing computation in scientific applications;
\item Identify the diverging effect of concurrency change on read and write in applications, and demonstrate the effectiveness of write-aware data placement. 
\item Develop a prediction model to estimate performance at various concurrency and data sizes to select optimal configurations.
\end{itemize}

\section{Background}\label{sec:bg}
In this section, we introduce NVM-based memory systems and the Seven Dwarfs in scientific applications. 

\subsection{NVM-based Heterogeneous Memory}
Extensive research has proposed using NVM for implementing the main memory to exploit its high density, persistence, and power efficiency~\cite{Qureshi2009,lee2009architecting}. Still, the current NVM technologies have lower performance than DRAM, and thus, main memory designs often pair NVM with DRAM, either as a cache or placed side-by-side to NVM. Previous works mostly use simulations and small problems for evaluation due to the lack of large-scale hardware. Recently, the first mass production of byte-addressable NVM arrives in the format of the Intel Optane DC Persistent Memory Module (PMM). In this work, we use this new hardware to evaluate realistic problems on promising memory designs.

\begin{figure}[bt]
	\centering
	\includegraphics[width=\linewidth]{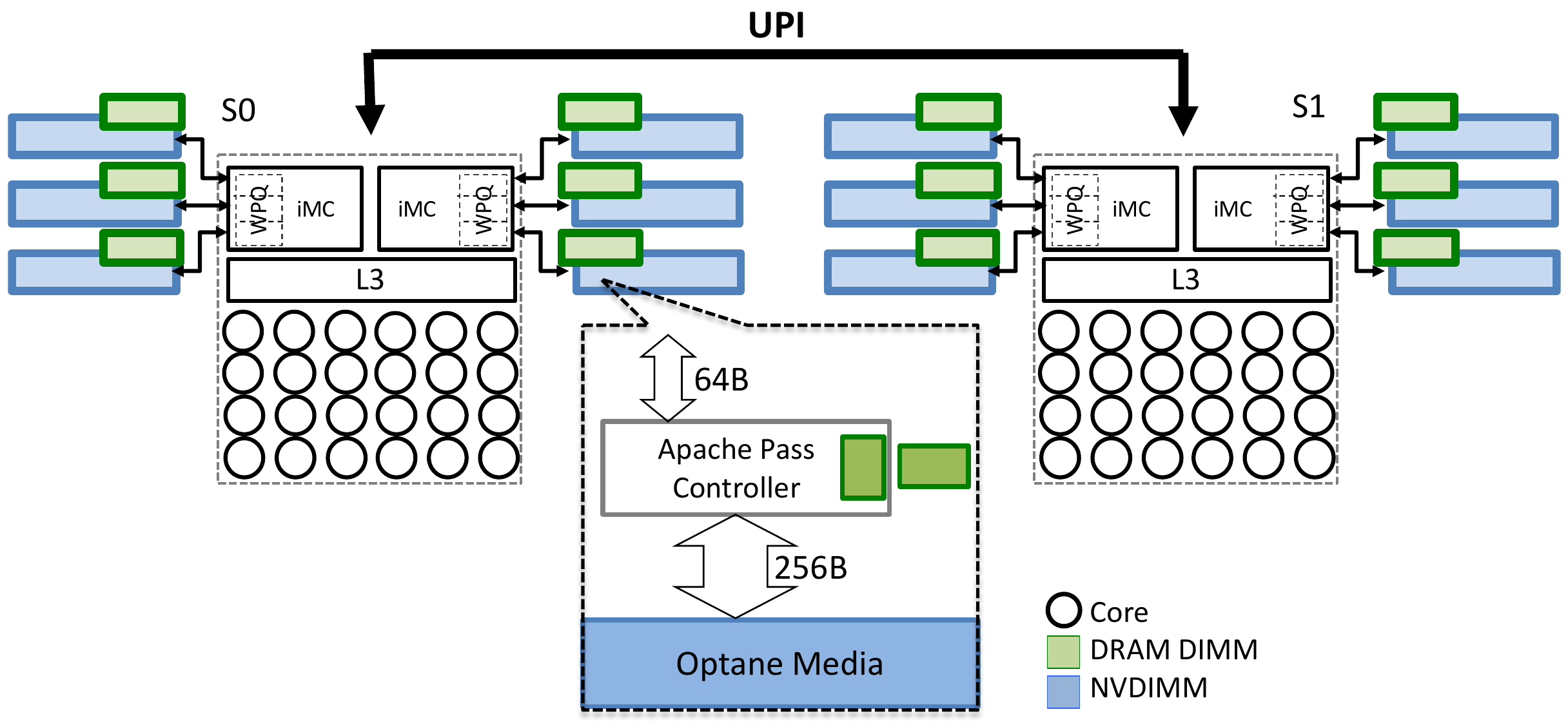}
	\caption{The system architecture the Intel Purley platform. \label{fig:arch}}
	
\end{figure}

The work of~\cite{peng2019} has provided detailed system evaluation, and we briefly summarize the system architecture (Figure~\ref{fig:arch}) in this section. The memory subsystem consists of DRAM DIMMs and NVDIMMs that share integrated memory controllers (iMC). Each NVDIMM has a small internal controller for address translation and a data buffer. The internal data granularity in the Optane media is $256$ bytes, while the data granularity between the processor and memory subsystem is $64$ bytes. System evaluation has quantified that sequential and random read accesses to NVM have a latency of $174$~ns and $304$~ns, respectively~\cite{peng2019}. Write latency to NVM depends on store instructions and data sizes. For instance, $64$- to $256$-byte non-temporal data store has $180$ -- $200$~ns latency~\cite{ucsd19}. On one socket, the read bandwidth to NVM can reach $39$~GB/s while the peak write bandwidth is only $13$~GB/s~\cite{peng2019,ucsd19}. Thus, the NVM exhibits about three times asymmetry in read and write bandwidth.

The NVDIMMs can be configured in \emph{Memory} or \emph{AppDirect} mode. In Memory mode, DRAM becomes a hardware-managed direct-mapped write-back cache to NVM and is transparent to applications. Note that DRAM on one socket cannot cache accesses to NVM on another socket~\cite{utexas19}. In AppDirect mode, the NVM becomes a byte-addressable persistent memory. A \emph{dax}-aware file system would transparently convert file read and write operations into $64$-byte load and store instructions in this mode to access NVM. Also, in this mode, the NVM on each socket can be exposed as a non-uniform memory access (NUMA) node to the CPUs. Standard NUMA management routines like \emph{numactl} can be used to control data placement in this configuration.

\subsection{Seven Dwarfs of HPC scientific applications}
The work of~\cite{asanovic2006landscape} summarizes seven domains of numerical algorithms in major HPC science and engineering applications, known as ``Seven Dwarfs". For a comprehensive evaluation of the HPC landscape, we select one application from each Dwarf as well as Laghos~\cite{dobrev2012high}, a proxy application of the BLAST hydrodynamics application, for the experiments. We introduce each Dwarf and application as follows. 
\begin{itemize}[leftmargin=*]
 \item \textbf{Dense Linear Algebra} features dense array data structures. They exhibit strided memory access to all the elements of the data structures. Classic vector and matrix operations fall into this category. We select matrix multiplication (level 3) from ScaLAPACK~\cite{blackford1997scalapack} for the experiment.  	
 \item \textbf{Sparse Linear Algebra} methods store data in compressed formats and access data elements through indirect memory accesses. We choose SuperLU~\cite{li2005overview} that adopts the BAR method for implementing sparse LU factorization.
 \item \textbf{Spectral Methods} often use fast Fourier transforms (FFT) to solve differential equations. Data permutation in this method often requires matrix transpose. We evaluate the FT benchmark that performs discrete 3D FFT from the NPB~\cite{bailey1991parallel} suite.
 \item \textbf{N-Body Methods} have a high computation complexity of $\mathcal{O}{(N^2)}$ for simulating a dynamical system of $N$ particles. We use hardware accelerated cosmology code (HACC) ~\cite{habib2013hacc} that simulates the formation of structure in collisionless fluids under the influence of gravity in an expanding universe.	
 \item \textbf{Structured Grids} feature regular grid structures. Stencil operations on the grids often have high spatial locality in data accesses. We choose Hypre~\cite{yang2006parallel}, a high-performance pre-conditioners library for solving linear systems in our evaluation.
\item \textbf{Unstructured Grids} feature irregular grid structures. Data accesses and updates often involve multiple levels of memory reference indirection. We use a general block-structured AMR framework, BoxLib~\cite{bell2012boxlib}, for the test.  
\item \textbf{Monte Carlo} methods rely on repeated random data accesses to calculate numerical results. We use XSBench~\cite{tramm2014xsbench}, which implements a Monte Carlo neutron transport algorithm, as a representative of such workloads. 	
\end{itemize}
\section{Methodology}\label{sec:setup}
In this section, we describe the experimental setup, benchmarks, and methodologies. We use the Intel Purley platform that consists of two 2\textsuperscript{nd} Gen Intel\registermark Xeon\registermark Scalable processors as the testbed. The memory subsystem consists of four iMCs, $12$ memory channels, a total of 192 GB DRAM ($12$ DIMMs), and 1.5 TB NVM ($12$ NVDIMMs). The memory channels run at 2400 GT/s, supporting $230.4$~GB/s peak system bandwidth. We override the EFI memory map to expose NVM on each socket as a separate NUMA node. The configuration of the system is summarized in Table~\ref{tab:setup}.

The platform runs the Fedora 29 operating system with GNU/Linux 5.1.0. When the Optane DC PMM is configured in AppDirect mode and exposed as NUMA nodes, we use \textit{numactl} to control the data placement onto different memories. Table~\ref{tab:app} summarizes the applications and their input problems. We compile all applications with GCC 8.3.1. For each application, we report the application-defined figure of metric (FoM) if available. Otherwise, we report the run time of the main computation kernels. 

We develop profiling routines that sample memory bandwidth on each NVDIMM and DRAM DIMM. We use the Intel Processor Counter Monitor (PCM) tool~\cite{pcm} to monitor hardware counters to collect core activities and offcore events. The profiling routines are integrated into applications to exclude the initialization and finalization stages. We only report the profiling results of the main computation phases. When the Optane is configured in Memory mode, we report the memory traffic as measured to DRAM DIMMs because it is accessed before NVDIMMs. When the Optane is in AppDirect mode, we report the memory traffic as the sum of traffic to NVDIMMs and DRAM DIMMs. 

\begin{table}[bt]
	\caption{Platform Specifications\label{tab:setup}}
	\center
	\makegapedcells
	\resizebox{\columnwidth}{!}{
		\begin{tabular}{|c||l|}
			\hline
			\textbf{Processor} 	& 2\textsuperscript{nd} Gen  Intel\registermark Xeon\registermark Scalable processor\\\hline
			\textbf{Cores}		& 2.4 GHz (3.9 GHz Turbo frequency $\times$ 24 cores (48 HT) $\times$ 2 sockets\\\hline
			\textbf{L1-icache} 	& private, 32 KB, 8-way set associative, write-back\\\hline
			\textbf{L1-dcache} 	& private, 32 KB, 8-way set associative, write-back\\\hline
			\textbf{L2-cache} 	& private, 1MB, 16-way set associative, write-back\\\hline
			\textbf{L3-Cache} 	& shared, 35.75 MB, 11-way set associative, non-inclusive write-back\\\hline
			\textbf{DRAM} 		& six 16-GB DDR4 DIMMs $\times$ 2 sockets\\ \hline
			\textbf{NVM} 		& six 128-GB Optane DC NVDIMMs $\times$ 2 sockets\\ \hline
			\textbf{Interconnect} 	& Intel\registermark UPI at 10.4 GT/s, 10.4GT/s, and 9.6 GT/s\\ \hline
	\end{tabular}}
\end{table}
\begin{table}[bt]
	\caption{Evaluated benchmarks.\label{tab:app}}
	\resizebox{\linewidth}{!}{
	\centering
	\begin{tabular}{|l|l|l|}
	\hline
	\textbf{Benchmark} 						&\textbf{Input Problems}\\\hline
	Hypre~\cite{yang2006parallel}				&a 3D electromagnetic diffusion problem\\
	Laghos~\cite{dobrev2012high}				&the Sedov blast wave Q3-Q2 3D computation\\
	ScaLAPACK~\cite{blackford1997scalapack}	&the distributed matrix multiplication of dimension NxN\\
	NPB~\cite{bailey1991parallel} - FT			&a discrete 3D fast Fourier Transform of class D\\
	HACC~\cite{habib2013hacc}				&a 252 simulation box using 384 grids in CORAL benchmark suite\\
	BoxLib (AMReX)~\cite{bell2012boxlib}		&the spherical chemical wave propagation\\
	XSBench~\cite{tramm2014xsbench}			&the unionized grid of XL problem with 34 million lookups\\
	SuperLU~\cite{li2005overview}				&a distributed PDGSSVX routine with real datasets from~\cite{davis2011university}\\ \hline		
	\end{tabular}}
\end{table}

\section{Performance Analysis}\label{sec:char}
This section evaluates the impact of NVM-based main memory on HPC scientific applications from three aspects -- (1) the performance sensitivity in cached-and uncached-NVM (2) the write throttling effect on critical phases (3) the diverging effect of concurrency changes on read and write accesses. We also quantify the performance impact of checkpointing on NVM. 

\subsection{Overall Performance}
\begin{figure*}
\centering
\includegraphics[width=\linewidth]{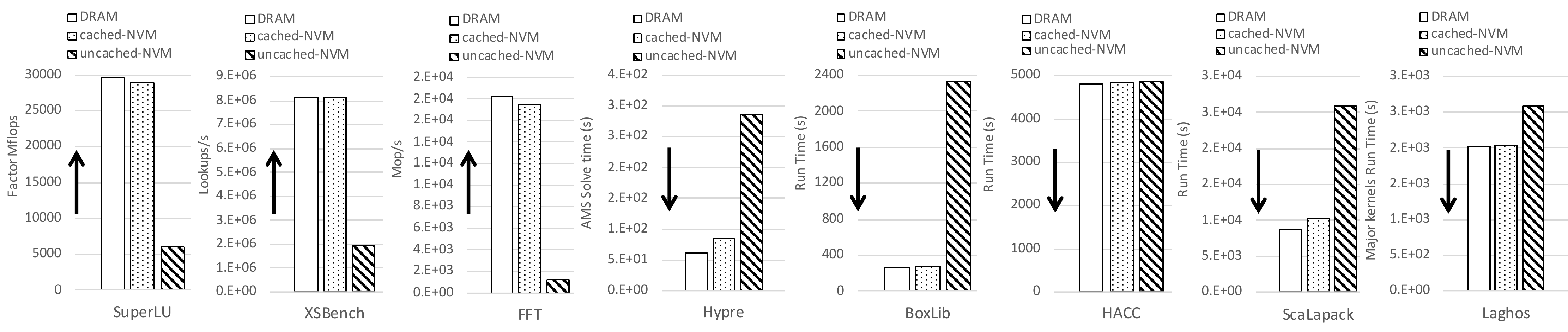}
\caption{An overview of the performance sensitivity of eight applications to cached and uncached NVM compared to DRAM. SuperLU, XSBench, and FFT report application-defined performance metrics and the others report the run time. \label{fig:alltime}}
\end{figure*}
\begin{table*}
\caption{An overall characterization of application sensitivity to the uncached-NVM. The last column indicates the performance compared to that on DRAM. Highlighted cells classify applications into three tiers.\label{tab:classify}}
\centering
\begin{tabular}{|l|c|c|c|c|c|c|}
\hline
\textbf{Dwarf} &\textbf{Application} &\textbf{Memory BW (MB/s)} &\textbf{Read BW (MB/s)} &\textbf{Write BW (MB/s)} &\textbf{Write Ratio($\%$)} &\textbf{Slowdown(x)} \\\hline\hline
N-body				&HACC		&40		&25.4	&14.3	&36	&\cellcolor{green!25}1.01\\
Structured Grid 		&Lagos		&4,135	&3,114	&1,021	&25  &\cellcolor{green!25}1.27\\
Dense Linear Algebra  	&Scalapack	&11,984	&10,104	&1,880	&16 &\cellcolor{gray!50}2.99\\
Monte Carlo  			&XSBench 	&16,134	&16,130	&4		&0	&\cellcolor{gray!50}4.16\\
Structured Grids 		&Hypre 		&11,413	&10,519	&894		&8    &\cellcolor{gray!50}4.67\\
Sparse Linear Algebra 	&SuperLU		&8,342	&6,208	&2,134	&25 &\cellcolor{gray!70}4.94\\
Unstructured Grids 		&BoxLib		&10,336	&8,248	&2,088	&21 &\cellcolor{blue!40}8.94\\
Spectral Methods		&FFT		&5,983	&3,633	&2,350	&39 &\cellcolor{blue!40}14.92\\
\hline
\end{tabular}
\end{table*}

We start the evaluation with an overview of the performance sensitivity of HPC applications on two NVM-based main memory, i.e., cached and uncached. All the experiments use the local socket to eliminate the severe NUMA effects reported in~\cite{utexas19,peng2019,ucsd19}. Input problems have a memory footprint fit in DRAM capacity ($50\%$-$85\%$) so that we can use the performance on DRAM-only main memory as the reference. Figure~\ref{fig:alltime} reports application performance on DRAM-only, cached-NVM, and uncached-NVM main memory. Note that SuperLU, XSBench, and FFT use application-defined metrics, i.e., the higher, the better, while the other applications report the run time, i.e., the lower, the better.

All applications on the cached-NVM manage to achieve performance comparable to that on the DRAM. The performance gap between DRAM and the cached-NVM is less than $10\%$ except for ScaLAPACK, Hyre, and BoxLib. These three applications have more performance loss, with a maximum loss of $28\%$ in Hypre (to be analyzed in Section~\ref{sec:cache}). Note that cached-NVM requires no porting efforts from the application developer, which would likely be the first deployment choice. 

On the uncached NVM, applications exhibit three tiers of performance sensitivity, i.e., \emph{insensitive}, \emph{scaled}, and \emph{bottlenecked} performance compared to the DRAM baseline. We report the profiling results of average memory traffic, in read and write, respectively, in Table~\ref{tab:classify}. In the first tier, HACC and Laghos show little performance change when the main memory changes from DRAM-based to NVM-based. The performance loss is much lower than the latency and bandwidth difference between NVM and DRAM. This class of applications features low memory bandwidth (in green). Scientific applications that share similar computations as HACC (N-body) and Laghos (unstructured finite element) may sustain performance when directly ported to systems with NVM-based main memory.  

Four applications (in gray cells) exhibit scaled performance on the uncached NVM, as compared to their DRAM baseline. These applications exhibit $2.99$ to $4.94$ times slowdown, which approximates the three times performance gap between DRAM and NVM, as benchmarked on the testbed~\cite{peng2019}. ScaLAPACK, XSBench, and Hypre have a high memory bandwidth but a low write ratio. The total memory bandwidth ranges between $11$ and $16$ GB/s while their write accesses only take up $0.03\%$ to $16\%$ total memory traffic. SuperLU falls between the scaled and bottlenecked tier because its main computation phase, ``factor'', consists of two dramatically different stages. The first phase slows down more than ten times, but the second phase has no performance loss. Section~\ref{sec:write} provides in-depth analysis.

The third tier of performance sensitivity, including BoxLib and FFT, shows a severe performance bottleneck on the uncached NVM. They slow down more than the performance gap between DRAM and NVM. This group of applications has a total memory bandwidth lower than applications in the second tier but high write traffic. In particular, their memory traffic has read/write ratios as low as $1.5$ so that the write traffic could even reach $36\%$ total memory traffic. We notice that the hardware supports up to  $12$~GB/s write bandwidth on the testbed, indicating that bandwidth saturation is not the cause of the bottleneck. We identify the write throttling effect and concurrency contention as the primary causes of the slowdown in Section~\ref{sec:contention}.

\noindent\emph{Insight I: N-body and structured grids applications may have negligible performance loss on NVM-based main memory.}

\subsection{Cache Efficiency}\label{sec:cache}
\begin{figure*}
     \centering     
     \begin{subfigure}{0.3\linewidth}
         \centering
         \includegraphics[width=\linewidth]{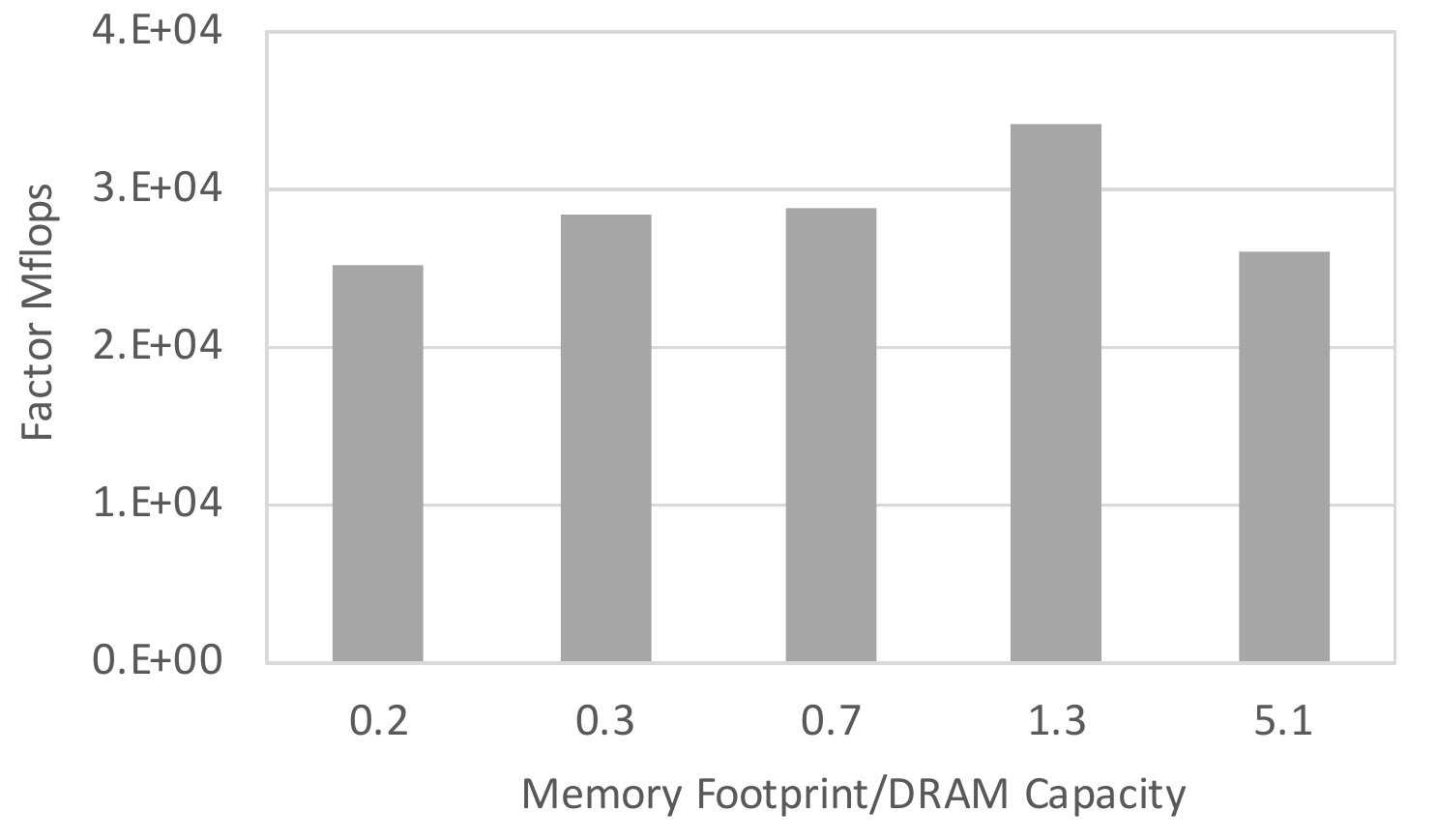}
         \caption{SuperLU. \label{fig:cache_lu}}
     \end{subfigure}
     ~
     \begin{subfigure}{0.3\linewidth}
         \centering
         \includegraphics[width=\linewidth]{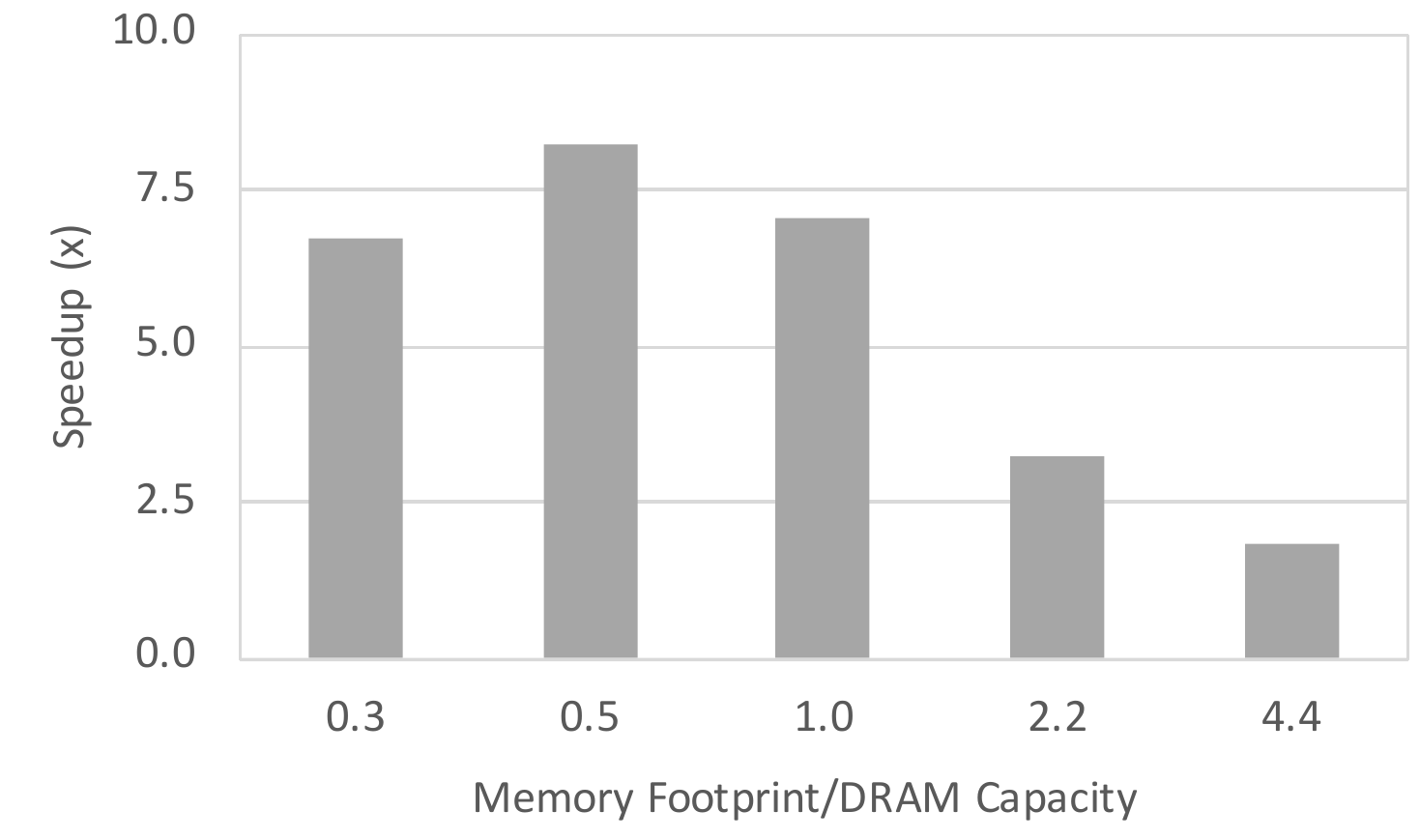}
         \caption{BoxLib. \label{fig:cache_boxlib}}
     \end{subfigure}
     ~
     \begin{subfigure}{0.3\linewidth}
         \centering
         \includegraphics[width=\linewidth]{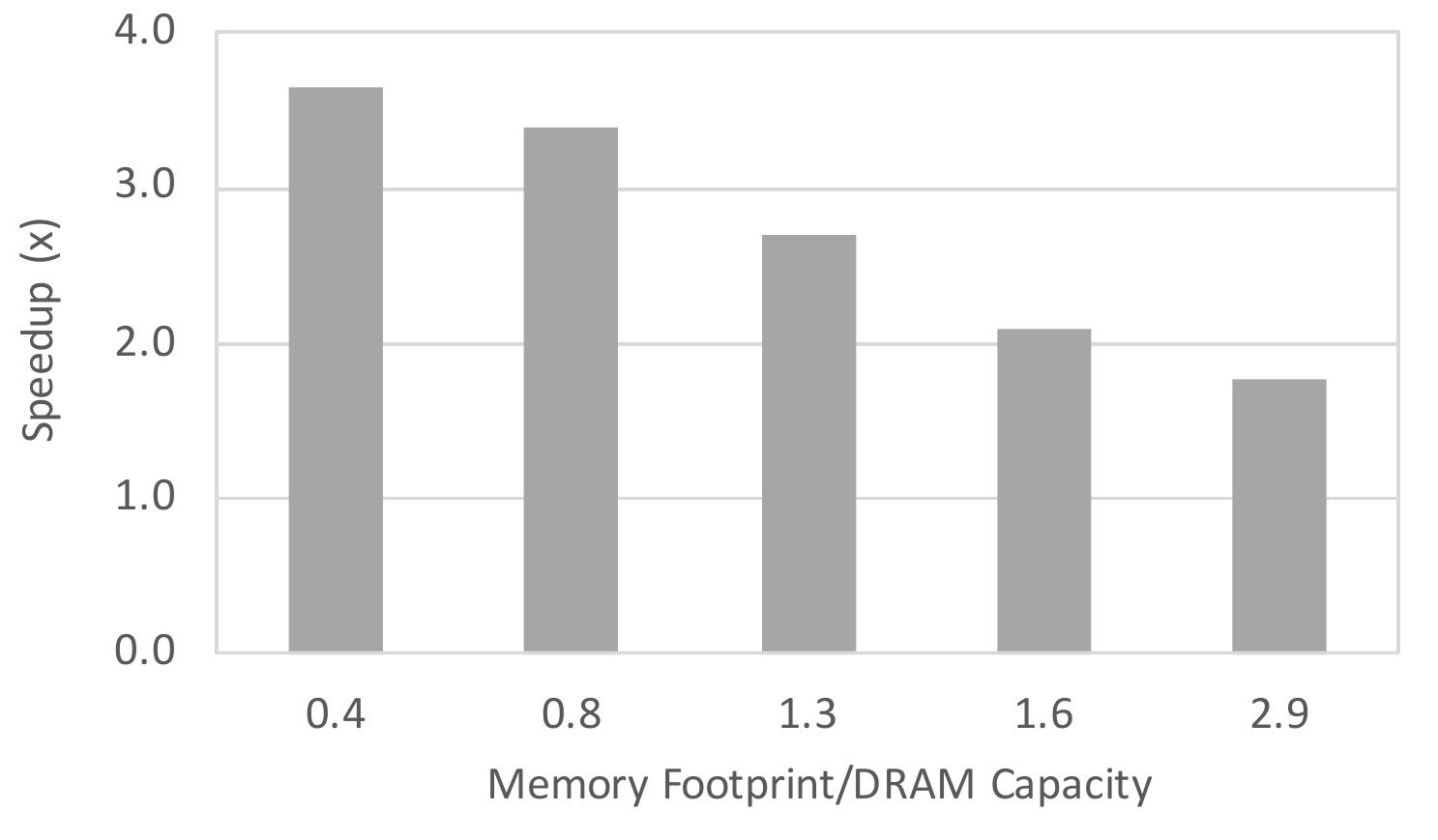}
         \caption{Hypre. \label{fig:cache_hypre}}
     \end{subfigure}
     \caption{The performance of SuperLU, BoxLib, and Hypre at increased input problems. SuperLU reports the application-defined metric. BoxLib and Hypre report the speedup on cached-NVM compared to noncached-NVM. \label{fig:cache2}}
 \end{figure*}
\begin{figure}[ht]
	\centering
	\includegraphics[width=\linewidth]{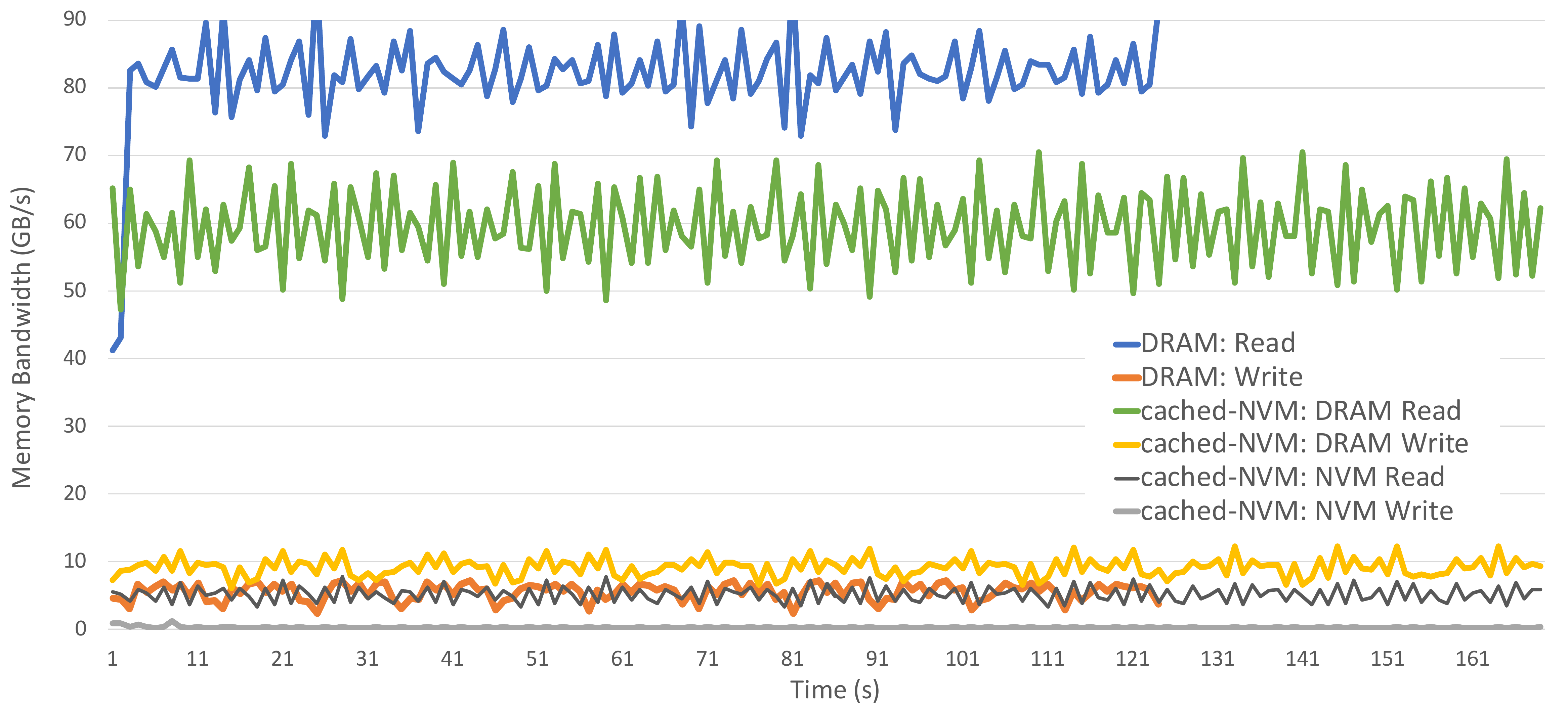}
	\caption{The trace of read and write bandwidth when Hypre run on cached-NVM and DRAM, respectively. \label{fig:cache1}}
\end{figure}

We use two metrics to quantify the effectiveness of using DRAM as a cached to NVM. First, for input problems with memory footprint smaller than the DRAM capacity, we define \emph{cache efficiency} as the relative performance to that on DRAM directly. The expectation is that low hardware overhead for managing DRAM as a cache should bring the performance to match using DRAM natively. Second, when the input problems require memory size larger than the DRAM capacity, we defined \emph{cached speedup} as the performance improvement from non-cached NVM. The expectation is that the cached-NVM can enable larger problems than DRAM and also higher performance than uncached-NVM.

Overall, the cached NVM delivers high cache efficiency. Hypre is the application with the most performance loss in the cached-NVM, i.e., $28\%$. We identify the cause by collecting samples of read and write traffic throughout the execution. In the cached-NVM, DRAM acts as the last level cache before the NVM main memory. Thus, we collect traffic to both DRAM and NVM. In DRAM-only main memory, we only collect traffic to DRAM DIMMs. Figure~\ref{fig:cache1} reports the reconstructed trace of memory traffic. We notice that the $59.5$~GB/s read bandwidth in cached-NVM (green line) is precisely $28\%$ reduction of the $82.5$~GB/s read bandwidth in DRAM (blue line). Since Hypre is a read-dominant workload, the read bandwidth directly affects the performance. Interestingly, the write bandwidth in cached-NVM (yellow line) reaches $9.3$~GB/s, which is significantly higher than the $5.7$~GB/s write bandwidth in DRAM (orange line). We attribute this increased write traffic to DRAM cache to the cache line replacement, i.e., load misses in the DRAM cache need to read from NVM (black line) and then save data into the DRAM cache. 

We scale up the input problems in three MPI applications, i.e., SuperLU, BoxLib, and Hypre, which typically require multiple compute nodes on supercomputers for realistic simulations. For SuperLU, we use five real datasets (kim2, offshore, Ge87H76, nlpkkt80, and nlpkkt120) from~\cite{davis2011university}. The largest input requires $490$~GB memory. For BoxLib and Hypre, we scale up their simulation domains to reach $300$~GB memory footprint. Figure~\ref{fig:cache2} reports the performance on the cached-NVM. SuperLU sustains similar performance (factor mflops) even when the input problems scale up to five times DRAM capacity. When BoxLib and Hypre have the memory footprint $4.4$ and $2.9$ times the DRAM capacity, the cached-NVM still manages to double the performance compared to the uncached-NVM. 

\noindent\emph{Insight II: sparse linear algebra and structured/unstructured grids applications can benefit from the cached-NVM to enable substantially large problems at reasonable performance.}

\subsection{Write Throttling}\label{sec:write}
\begin{figure*}
    \centering
    \begin{subfigure}{0.48\linewidth}
    \centering
    \includegraphics[width=\textwidth]{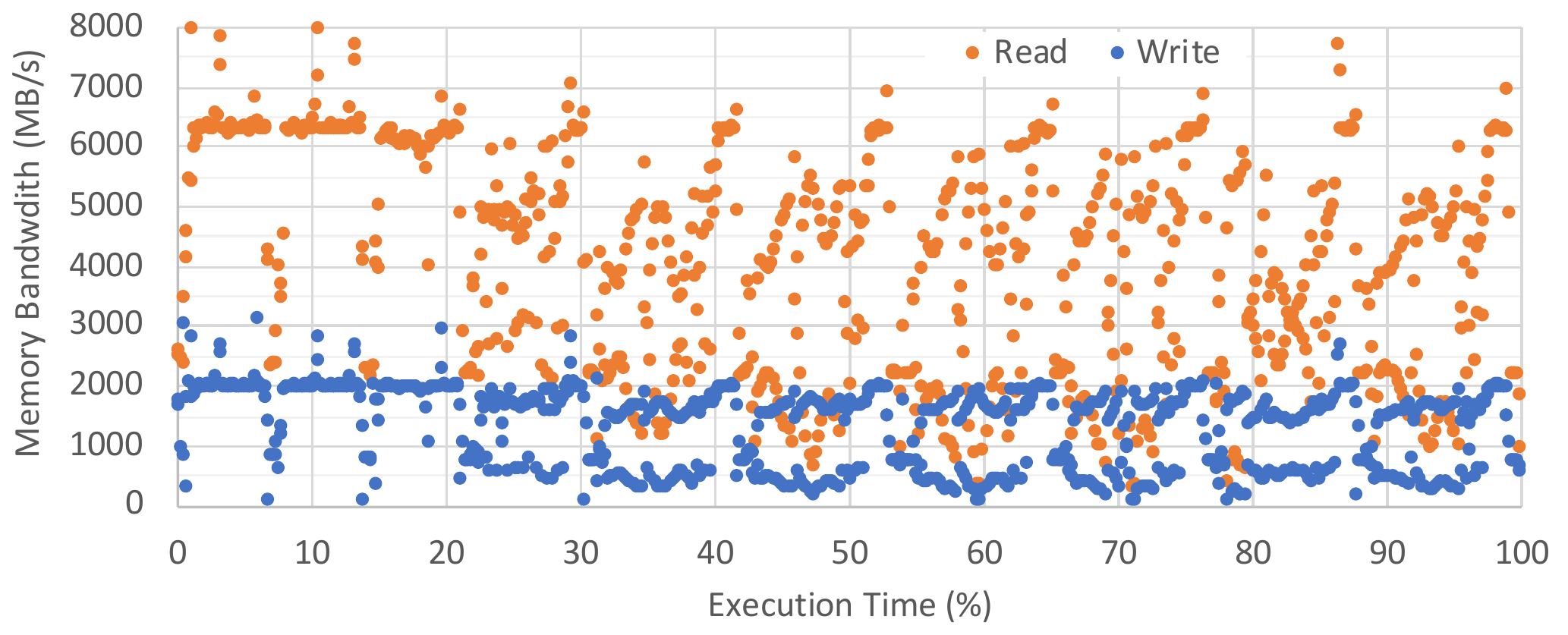}
    \caption{Laghos on DRAM\label{fig:bw_lag1}}
    \end{subfigure}
     ~
    \begin{subfigure}{0.48\linewidth}
         \centering
         \includegraphics[width=\textwidth]{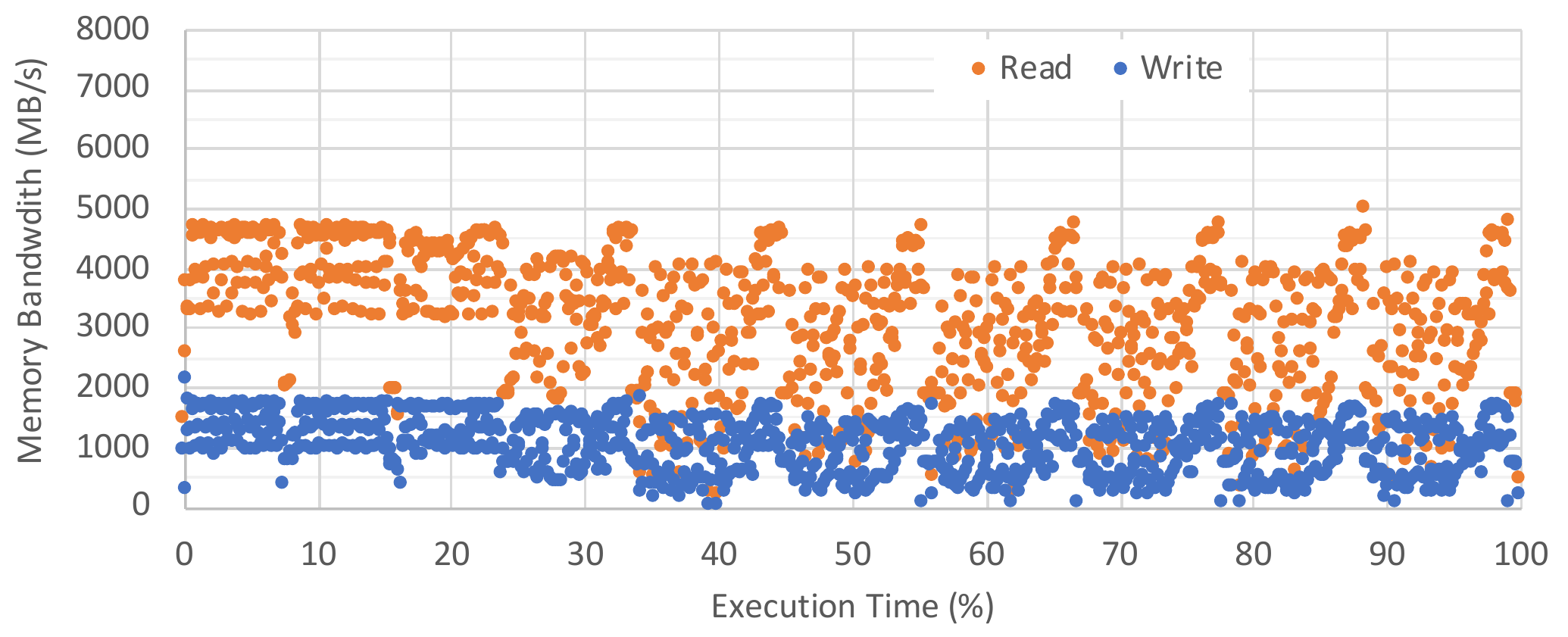}
         \caption{Laghos on uncached-NVM\label{fig:bw_lag2}}
     \end{subfigure}
         
    \begin{subfigure}{0.48\linewidth}
    \centering
    \includegraphics[width=\textwidth]{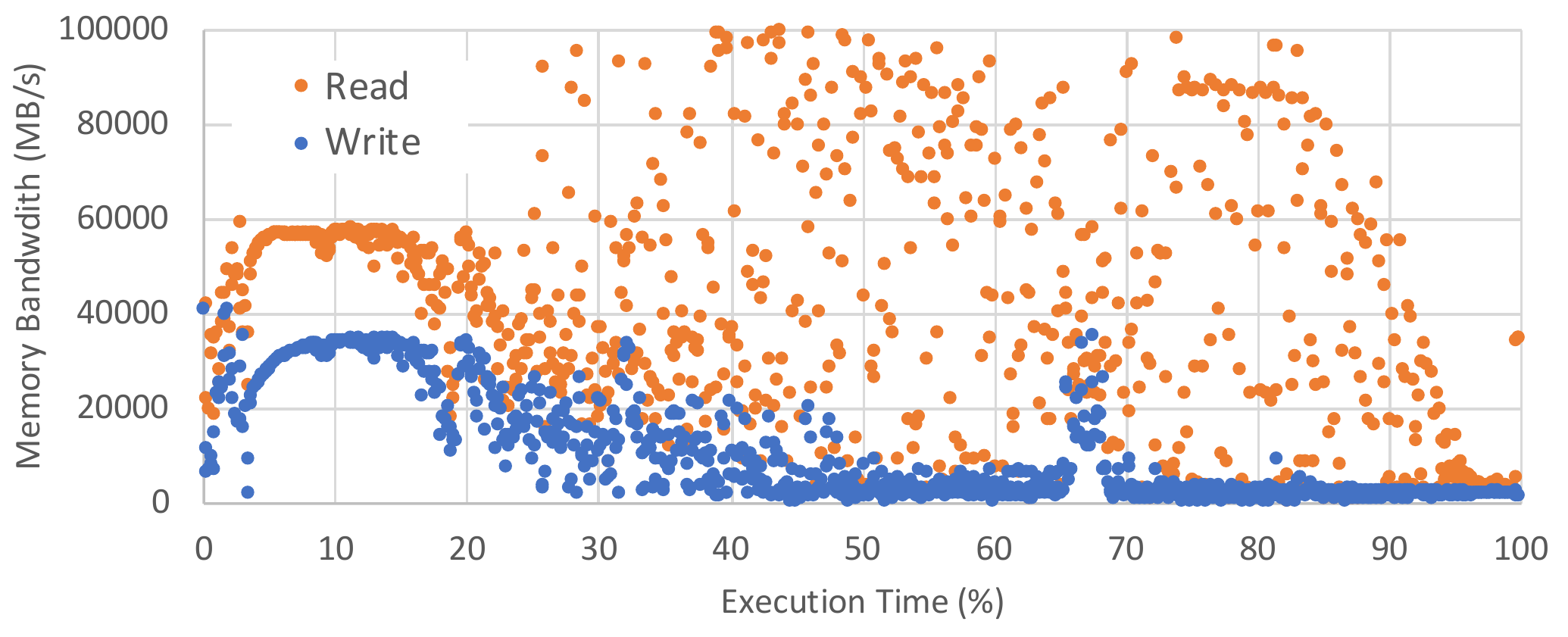}
    \caption{SuperLU on DRAM\label{fig:bw_lu1}}
    \end{subfigure}
     ~
    \begin{subfigure}{0.48\linewidth}
         \centering
         \includegraphics[width=\textwidth]{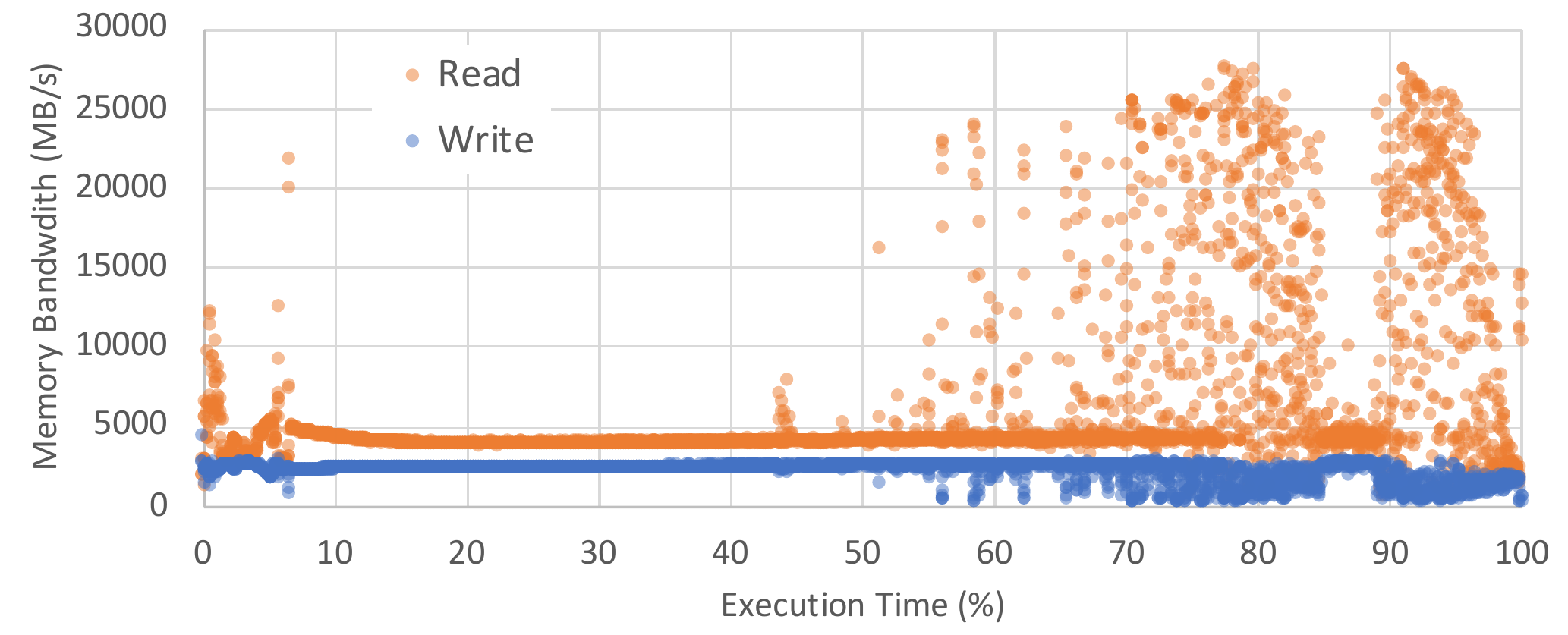}
         \caption{SuperLU on uncached-NVM\label{fig:bw_lu2}}
     \end{subfigure}
    \caption{Write throttling changes the dominant computation phase in SuperLU from 20\% to 70\% execution time. In contrast, Laghos sustains similar composition of computation phases on the uncached-NVM.\label{fig:trace}}
\end{figure*}

Uncached-NVM exposes the characteristics of NVM directly to the application without interference from the DRAM cache. We analyze application performance in this mode to provide feedback and insights for future NVM-based designs. Asymmetric read and write performance is a common characteristic of NVM technologies. For instance, the testbed in this study has $39$~GB/s read bandwidth and $13$~GB/s write bandwidth~\cite{peng2019}. Our analysis reveals a \emph{write-throttling} effect that could change the critical computation phase of an HPC application when moved from DRAM-based to NVM-based main memory. 

SuperLU and Laghos both have two distinct phases in the execution, as shown in the trace of memory traffic in Figure~\ref{fig:trace}. These phases, however, exhibit different sensitivity to the write throttling effect. The first phase in Laghos always takes about $20\%$ the execution time when running on DRAM (Figure~\ref{fig:bw_lag1}) and uncached-NVM (Figure~\ref{fig:bw_lag2}). In contrast, on DRAM, the first phase in SuperLU only takes $20\%$ the execution time (Figure~\ref{fig:bw_lu1}), but significantly extends to $70\%$ execution time on uncached-NVM (Figure~\ref{fig:bw_lu2}). 

We find that there exists a threshold value of the write bandwidth ($2$~GB/s on the testbed), above which a computation phase will significantly prolong the execution. For instance, the first phase in Laghos has a moving average of $1.3$~GB/s write bandwidth with its peak remaining lower than $2$~GB/s on DRAM. The read/write ratio remains at $3$ in this stage. These characteristics remain unchanged when Laghos runs on the uncached NVM. On the other hand, the first phase in SuperLU exhibits high write traffic, and low read/write ratio when running on DRAM, resulting in an average of $33$~GB/s and a peak at $40$~GB/s. When running on the uncached NVM, the write bandwidth of this phase reduced by about $14$ times, reaching only $2.3$~GB/s. The dramatically reduced write performance \emph{throttles} the read performance due to data dependency and coupling effects in shared units~\cite{peng2018siena}. Consequently, the read performance is also reduced significantly from $54$~GB/s to $4$~GB/s. It is a high priority to address this change of behavior in critical phases when optimizing applications on NVM-based main memory. 

We identify low read/write ratio and high write bandwidth as the indicator to detect applications that are susceptible to the write throttling effect. Comparing the two phases in SuperLU, we find that the second phase with a high read/write ratio and low write traffic has only a moderate slowdown, which is expected due to the performance gap between DRAM and NVM hardware. In Laghos, read and write bandwidth on DRAM in the two phases is lower than the peak bandwidth of NVM, so the changes in application performance are insignificant ($27\%$). We highlight that phase-specific characteristics become crucial for determining the bottleneck of HPC applications on NVM because the throttling effect could dramatically change the profile of execution. 

\noindent\emph{Insight III: HPC applications with computation phases susceptible to the write throttling effect on uncached-NVM require different priorities in optimization. }

\begin{figure}[bt]
	\centering
	\includegraphics[width=0.9\linewidth]{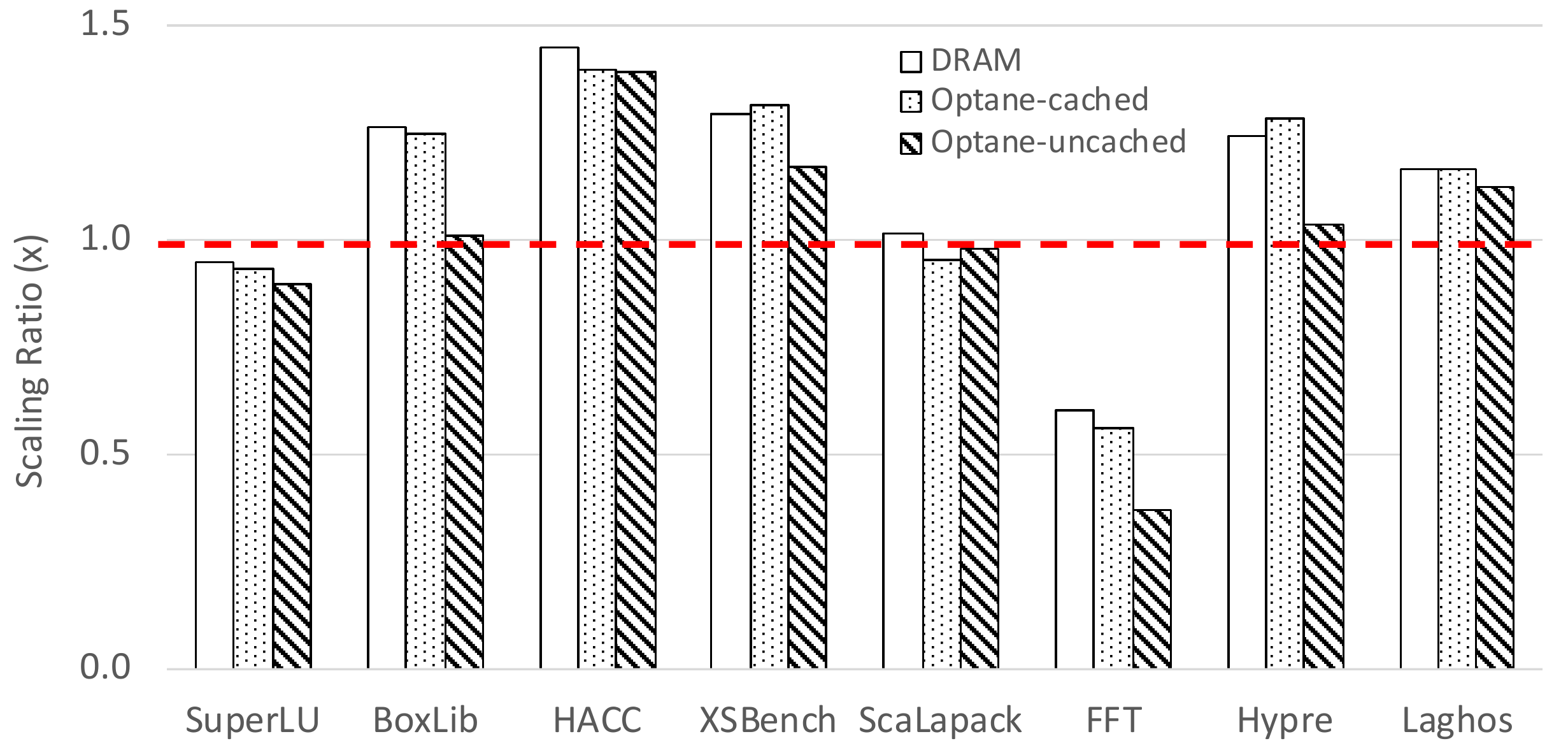}
	\caption{Ratio below the red line indicates performance loss when concurrency increases. Larger gap between DRAM and uncached-Optane indicates contention on NVM.\label{fig:contention}}
\end{figure}

\subsection{Concurrency Contention}\label{sec:contention}
\begin{figure*}
    \centering
    \begin{subfigure}{0.48\linewidth}
         \centering
         \includegraphics[width=\textwidth]{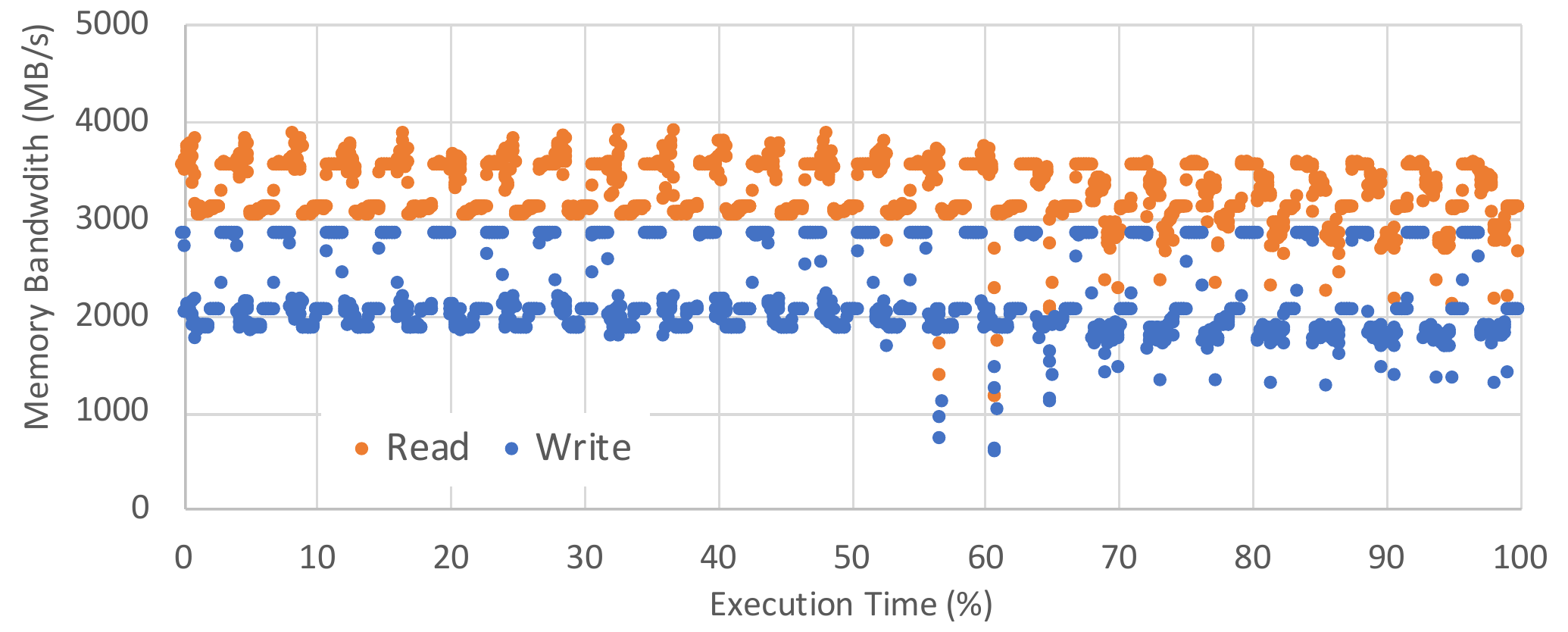}
         \caption{$Concurrency=8$\label{fig:ft1}}
     \end{subfigure}
 ~    
    \begin{subfigure}{0.48\linewidth}
    \centering
    \includegraphics[width=\textwidth]{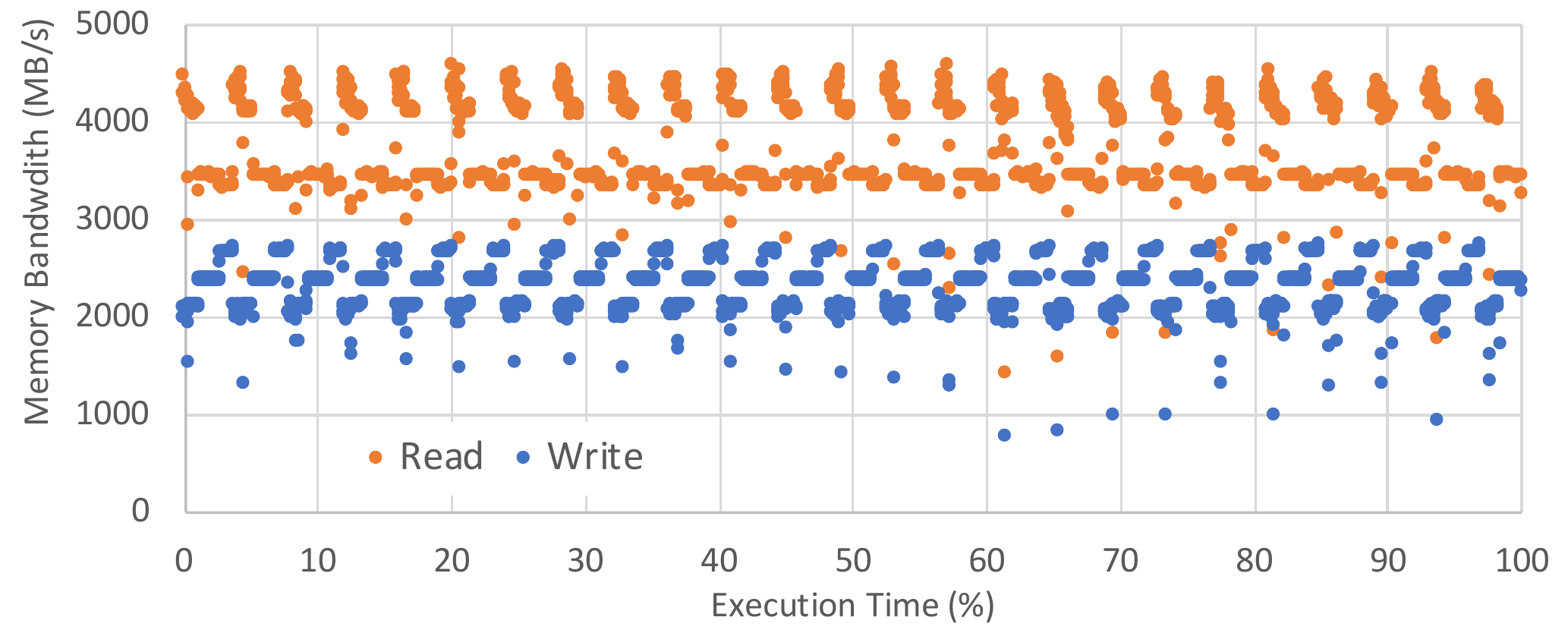}
    \caption{$Concurrency=24$\label{fig:ft2}}
    \end{subfigure}
    \caption{A diverging effect from concurrency changes read and write access in FT. The reduced write bandwidth overpowers the increased read bandwidth, resulting a $26\%$ performance loss.\label{fig:ft3}}
\end{figure*}
\begin{figure*}
    \centering
    \begin{subfigure}{0.48\linewidth}
         \centering
         \includegraphics[width=\textwidth]{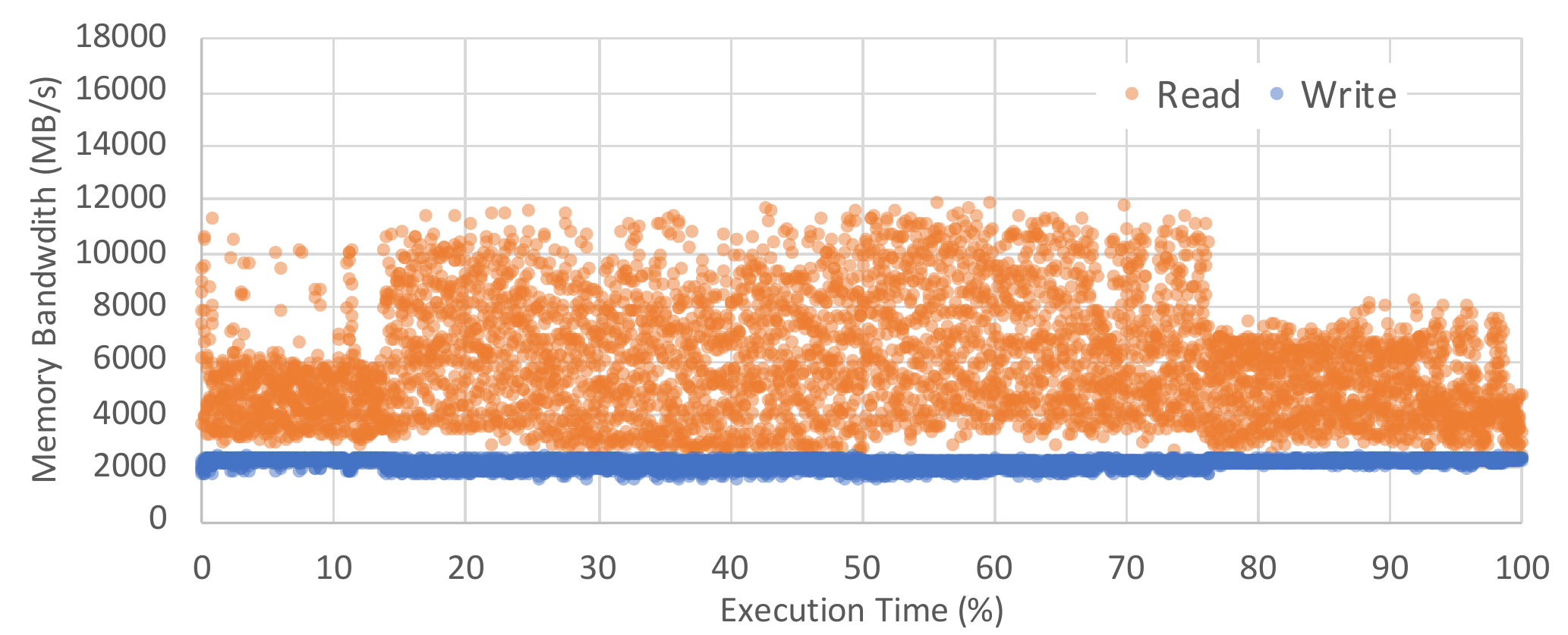}
         \caption{$Concurrency=16$\label{fig:scala1}}
     \end{subfigure}
~
    \begin{subfigure}{0.48\linewidth}
    \centering
    \includegraphics[width=\textwidth]{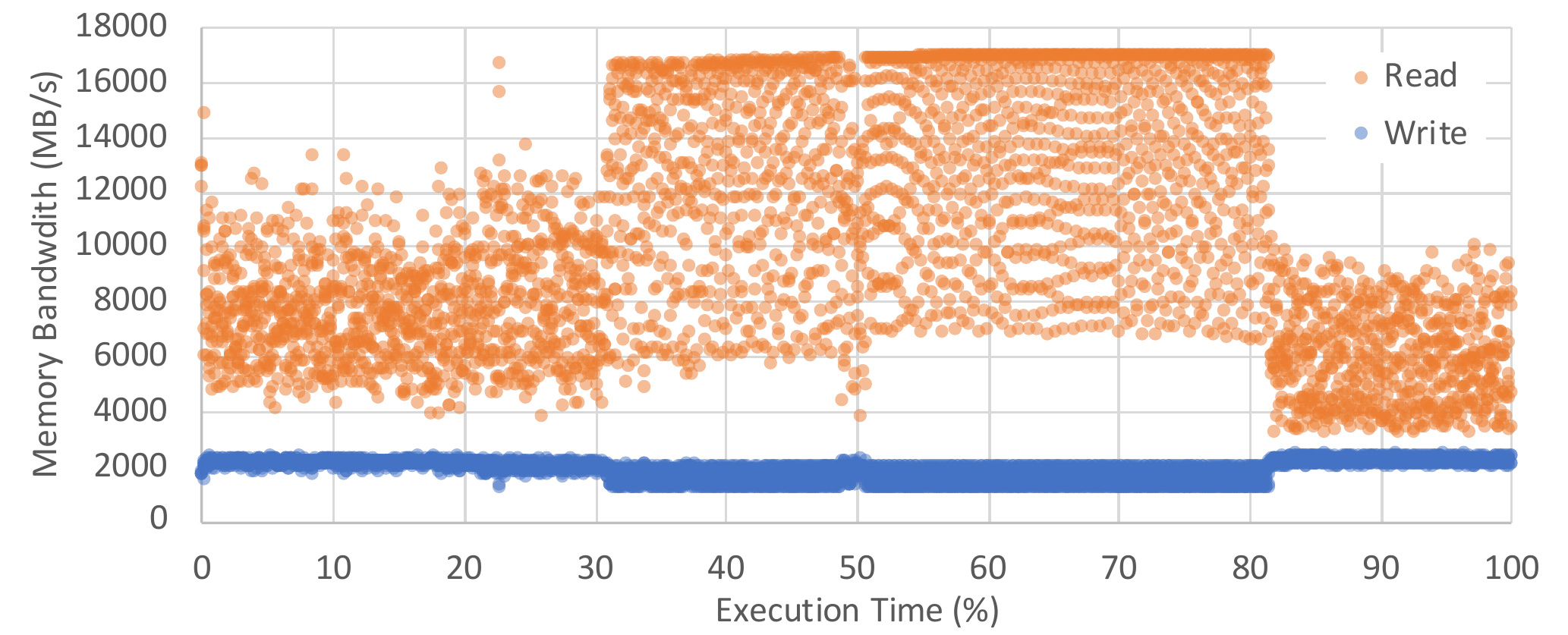}
    \caption{$Concurrency=36$\label{fig:scala2}}
    \end{subfigure}
    \caption{Increased concurrency in ScaLAPACK prolongs the first stage from 10\% execution time to 30\%, and also increases the gap between read and write bandwidth in the first $80\%$ execution time.\label{fig:scala3}}
\end{figure*}

HPC applications on supercomputers exploit the high parallelism from multicore processors to accelerate simulations. However, multiple threads may contend on shared buffers or units in the memory, creating a performance bottleneck. For instance, re-ordering and merging write to NVM is a common technique to mitigate the high energy cost and low write bandwidth of NVM technologies. On the testbed, write pending queues (WPQ) in NVM are used to combine multiple write requests into one transaction to the Optane media. When the concurrency is high, contention may arise on a fully occupied WPQ, where new requests have to wait for the WPQ to drain before being inserted into the queue~\cite{Yang2019AnEG}. Also, high concurrency would decrease the opportunity of combinable requests in WPQ, similar to the well-known fact that high concurrency reduces the locality in the row buffer. 

We identify concurrency contention on NVM by comparing the performance changes at different concurrency across memory configurations. We run each application using two levels of concurrency on DRAM, cached-NVM, and uncached-NVM, respectively. The \emph{contention ratio} on a memory configuration is measured as the performance at the high concurrency level normalized to that at the low concurrency level. Figure~\ref{fig:contention} reports the ratios in the eight applications. Applications with a ratio larger than one (the red dotted line) have performance improvement at increased concurrency. For instance, HACC and XSBench have more than $30\%$ performance improvement when their concurrency increases. A ratio below one (the red line) indicates performance loss at high concurrency. However, the loss may not necessarily be a result of contention on memory. Some algorithmic properties of low scalability could also cause performance loss. Therefore, we propose to compare ratios on NVM to that on DRAM to identify the concurrency contention. For instance, FFT has a ratio of $0.61$ on DRAM but only $0.37$ on uncached-NVM, indicating that the contention from NVM is the main cause for performance loss. Similarly, Boxlib has a notable gap between the two ratios. Interestingly, ScaLAPACK has higher contention on cached-NVM than uncached-NVM.

We reconstruct the trace of memory traffic of FT and ScaLAPACK at two concurrency levels in Figure~\ref{fig:ft3} and~\ref{fig:scala3}. FT consists of iterative phases. In each phase, the write bandwidth at the lower concurrency can reach $3$~GB/s (~\ref{fig:ft1}) while at the high concurrency it is below $2.6$~GB/s. The increased concurrency, however, has an opposite effect on read bandwidth, which increases from $3.8$~GB/s to $4.5$~GB/s. Overall, increased concurrency increases the divergence between the read and write bandwidth. This diverging effect could also change the composition of computation phases. In ScaLAPACK, the first stage extends from $10\%$ execution time in Figure~\ref{fig:scala1} to $30\%$ in Figure~\ref{fig:scala2}. Note that read bandwidth in the second stage increases from $12$~GB/s to $17$~GB/s, resulting in reduced execution time. Since the execution time of the first stage remains unchanged, it now becomes a more important phase in the computation. 

\noindent\emph{Insight IV: Concurrency changes may have a diverging effect on read and write access. Phase-specific optimization or write-aware data placement may be more effective than a global adjustment of concurrency.}

\subsection{Leveraging the memory persistence}
\begin{figure}[bt]
	\centering    
    \begin{subfigure}{0.45\linewidth}
    \centering
    \includegraphics[width=\textwidth]{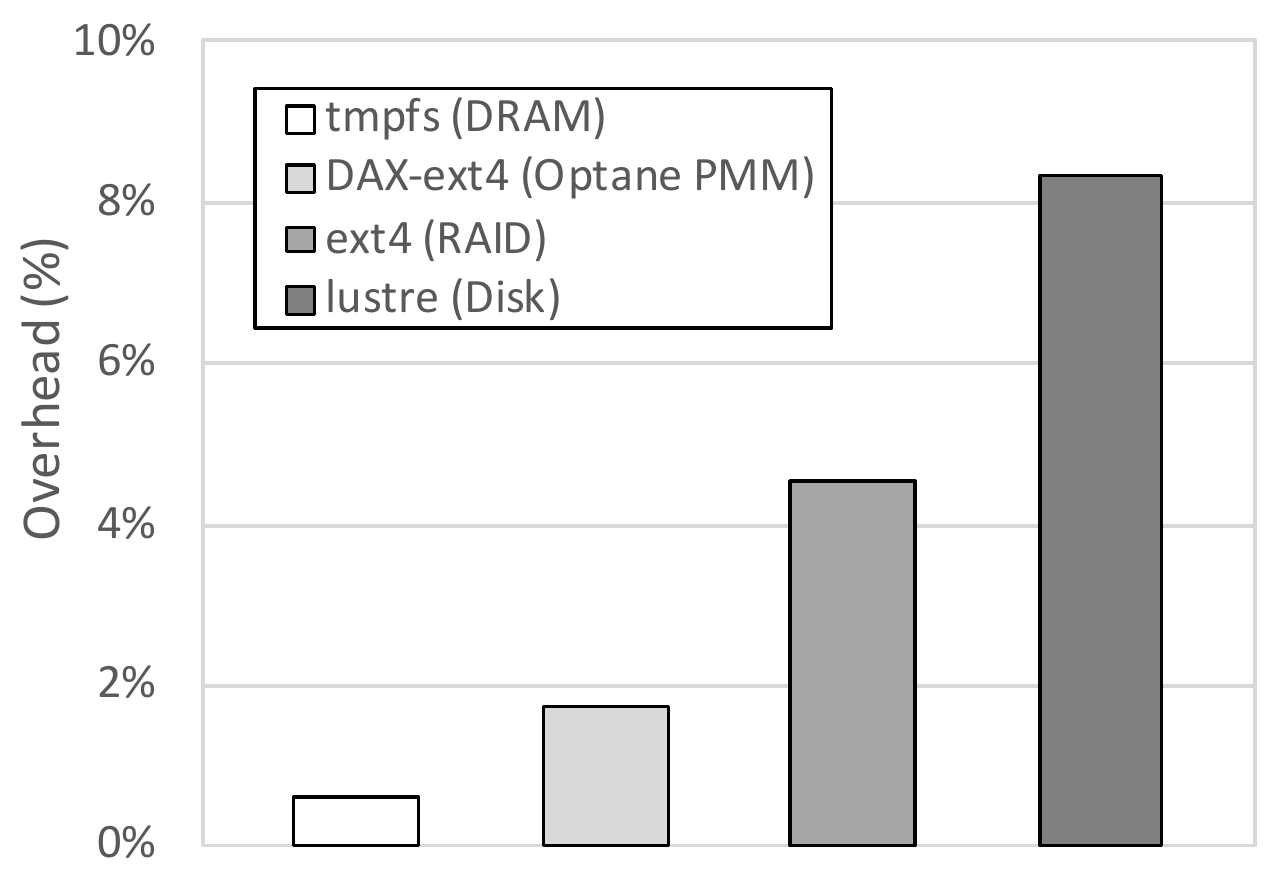}
    \caption{Overhead on four tiers of storage.\label{fig:checkpoint}}
    \end{subfigure}
    \begin{subfigure}{0.52\linewidth}
    \centering
    \includegraphics[width=\textwidth]{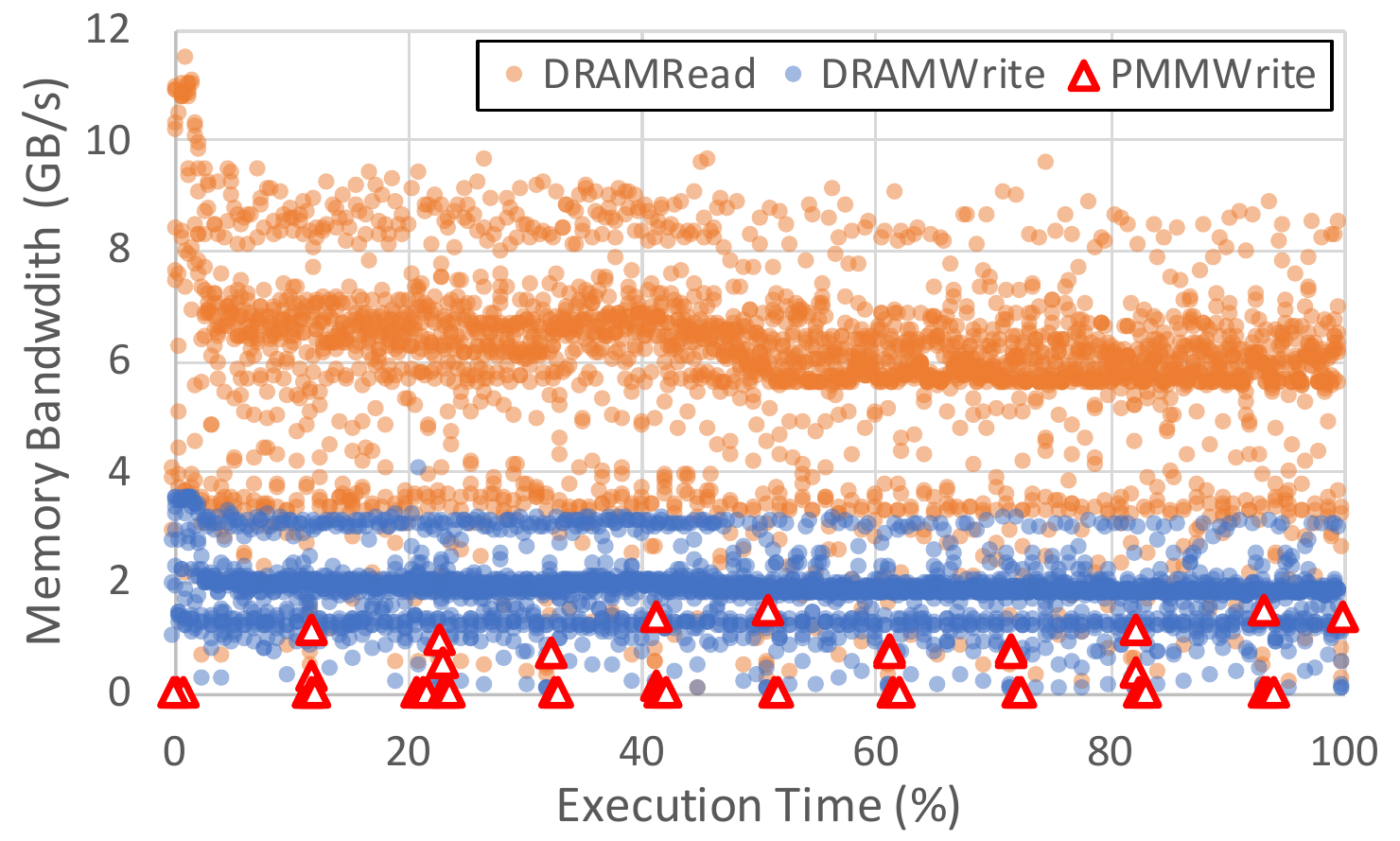}
    \caption{The trace of bandwidth.\label{fig:checkpoint2}}
    \end{subfigure}
    \caption{Snapshots in Laghos every five steps. \label{fig:checkpoint3}}
\end{figure}

Large-scale HPC simulations rely on I/O intensive visualization and checkpointing to detect anomaly at an early stage for long-running jobs. Thus, HPC applications may directly benefit from high bandwidth and persistence on Optane memory for I/O. We configure Optane in App Direct mode and evaluate the overhead of visualization in Laghos on four tiers of storage, from tmpfs on DRAM, a DAX-aware ext4 file system on the Optane, an ext4 file system on the local RAID, to a Lustre file system on network interconnected disk. Note that tmpfs is not persistent but provides the upper bound of performance. The results are consistent with the memory/storage hierarchy, as shown in Figure~\ref{fig:checkpoint}. The Optane memory only imposes $2\%$-$5\%$ overhead, achieving four times reduction in overhead on other persistent storages. We further analyze the interaction between NVM and DRAM traffic in Figure~\ref{fig:checkpoint2}. The write-only traffic to NVM is periodic (red triangles) at about 2~GB/s and shows no interference to the traffic to DRAM during the execution.

\section{Performance Optimization}\label{sec:opt}
In this section, we propose two optimization techniques for cached- and uncached-NVM main memory, respectively. We develop a prediction model to estimate performance in cached-NVM when the concurrency or data size changes. On uncached-NVM, we employ write-aware data placement to avoid the diverging effect.

\subsection{Model-based Prediction}
The performance analysis on cached-NVM in Section~\ref{sec:cache} and~\ref{sec:contention} shows that both concurrency change and data size can impact the effectiveness of execution. Naturally, if a performance model can predict application performance at various configurations, it helps the application developer select an optimal setup without exhaustively search the configuration space. Suppose a general configuration consists of multiple dimensions of freedom. Our empirical observation indicates that when sweeping the configuration in one dimension, performance impact may change from positive to negative, which will reflect in a similar trend on some hardware events. These events are defined as critical events and used as predictors~\cite{curtis2008prediction}. 

A set of critical events mutually indicate the overall trend in the performance of an application at a specific configuration. This behavior is modeled analytically in a multivariate function (Eq.~\ref{eq:model}). Here, each variant $N_{e_i}$ represents the count of one critical event $e_i$. $\beta_i$, the coefficient, indicates either a positive or negative impact on the derivative of performance. Our selection of critical events combines empirical observation and a statistical procedure. First, from the classification of performance sensitivity to NVM main memory, we identify several critical indicators, such as the computation intensity, memory traffic, read and write ratios. These metrics could be reflected in a range of hardware events. Second, we test a set of relevant hardware events into the regression model to prune highly correlated events, i.e., high p-values. Table~\ref{tab:events} summarize the events selected for deriving the prediction model. 
\begin{equation}
\label{eq:model}
IPC_{p} = \sum_{n=1}^{N} \beta_i \cdot (N_{e_i} \cdot IPC_{s}) + \sigma
\end{equation}
\begin{table}[ht]
	\caption{The events selected for performance prediction.\label{tab:events}}
	\centering
	\begin{tabular}{|c|l|l|}
	\hline
	\textbf{Feature}	&\textbf{Activities}\\\hline
	$p_0$	&Instruction Retired\\
	$p_1$	&Cycles Active\\
	$p_2$	&Cycles stalled due to Resource Related reason\\
	$p_3$	&Cycles in waiting for outstanding offcore requests\\
	$p_4$	&Count of the number of reads issued to memory controllers.\\
	$p_5$	&Counts of Writes Issued to the iMC by the HA.\\\hline
	\end{tabular}
\end{table}

We use two data collection strategies to collect training data sets for the models of concurrency and data size, separately. When deriving the model for predicting performance at different concurrency, we collect hardware events from application executions at the \emph{middle point} concurrency. For instance, for hardware with $HT$ hardware concurrency, we collect data sets from executions using $0.75 HT$. When deriving the model for predicting performance at a different data size, we fix the concurrency and collect events from configurations at a small data size. Next, the measurement for each hard event is first scaled by the sampled IPC ($IPC_s$ in Eq.~\ref{eq:model}) and then normalized by calculating their zero scores. The normalized features are used as the training data set to derive the coefficients of Eq.~\ref{eq:model} using multivariate linear regression. 

\begin{figure}[ht]
    \centering
    \begin{subfigure}{0.42\linewidth}
         \centering
         \includegraphics[width=\textwidth]{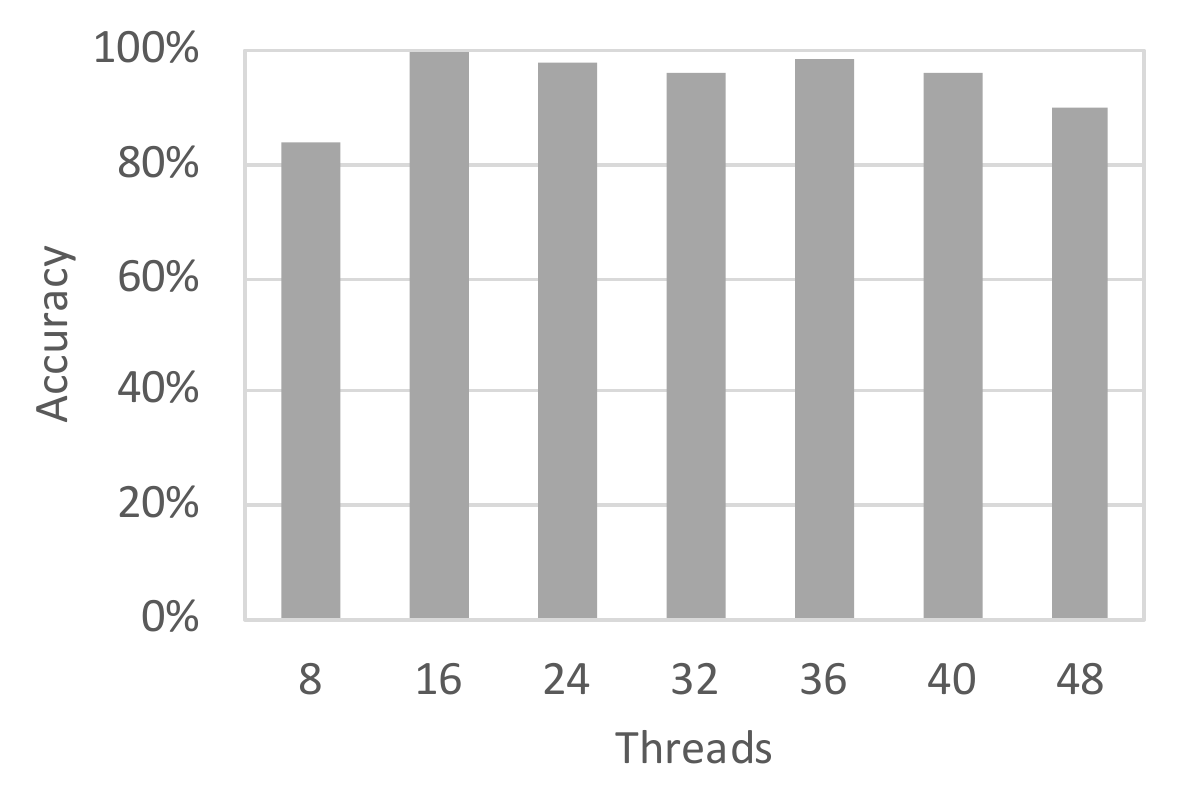}
         \caption{$XSBench$\label{fig:model1}}
     \end{subfigure}
     ~
    \begin{subfigure}{0.42\linewidth}
    \centering
    \includegraphics[width=\textwidth]{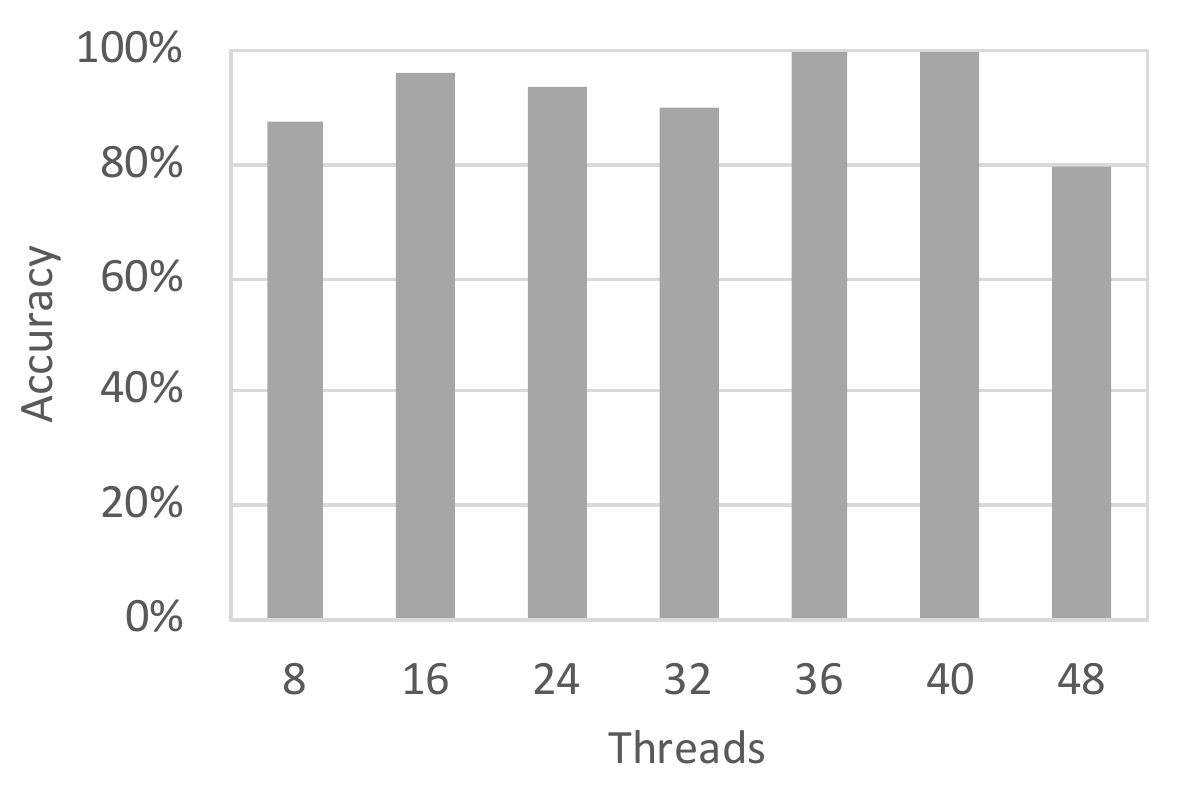}
    \caption{$NPB-FT$\label{fig:model2}}
    \end{subfigure}
    \caption{The model accuracy of concurrency change.\label{fig:model_ht}}
    \vspace{-15pt}
\end{figure}

\begin{figure}[ht]
    \centering
    \begin{subfigure}{0.38\linewidth}
         \centering
         \includegraphics[width=\textwidth]{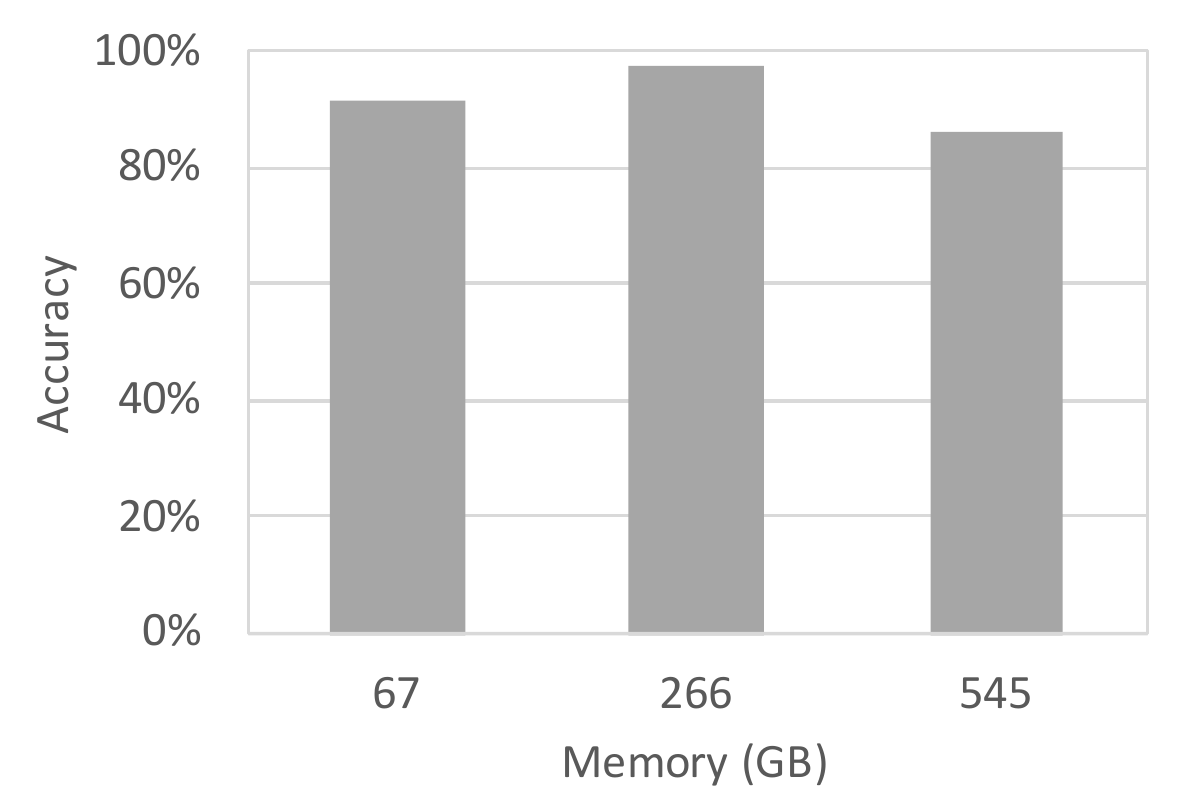}
         \caption{$XSBench$\label{fig:model3}}
     \end{subfigure}
     ~
    \begin{subfigure}{0.38\linewidth}
    \centering
    \includegraphics[width=\textwidth]{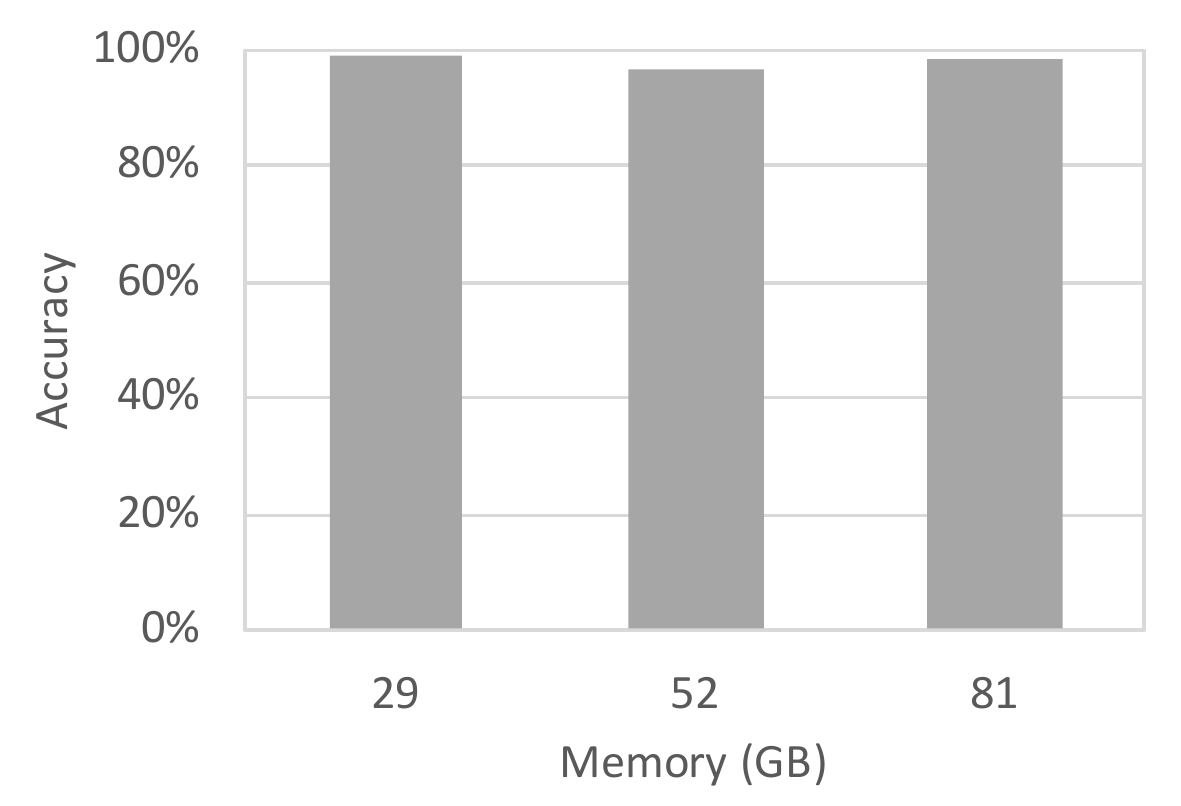}
    \caption{$ScaLAPACK$\label{fig:model4}}
    \end{subfigure}
    \caption{The model accuracy of data size change.\label{fig:model_size}}
\end{figure}

We evaluate the accuracy of the prediction by comparing the estimated IPC with the observed IPC. In the first experiment, we predict the application performance at different concurrency. We collect training data sets from running applications in the configuration of $ht=36$ only. The prediction model is derived from this training data set to estimate the performance at other concurrency levels. The estimation error $E_{est}$ is calculated as the absolute difference of prediction and observed divided by the observed IPC. In Figure~\ref{fig:model_ht}, we report the accuracy as $1-E_{est}$ for XSBench and FT, respectively. The average prediction errors in the two applications are $5\%$ and $8\%$. All concurrency levels except the lowest and highest level have accuracy above $90\%$. In the second experiment, we derive the prediction for various data sizes at concurrency $ht=36$. We collect the training data set from each application execution using a small input problem. The derived model then predicts the application performance at larger input problems. Figure~\ref{fig:model_size} reports the prediction accuracy when the data size (x-axis) increases in XSBench and ScaLAPACK. While all data sizes in ScaLAPACK have accuracy over $97\%$, XSBench has lower accuracy at the largest data size.


\subsection{Write-aware Data Placement}
We target uncached-NVM in heterogeneous memory in the second optimization strategy. Performance analysis in Section~\ref{sec:contention} has revealed that some applications, such as ScaLAPACK and FT, are sensitive to the diverging effect on read and write access from concurrency changes. Instead of adjusting the global concurrency level, we explore write-aware data placement in applications to circumvent this effect. This approach keeps data structures with substantial write traffic in DRAM and places other data structures in NVM. Therefore, when concurrency increases, the read bandwidth from NVM could scale up and write to DRRAM avoids the contention on NVDIMMs. 

\begin{figure}[ht]
	\centering
	\includegraphics[width=0.8\linewidth]{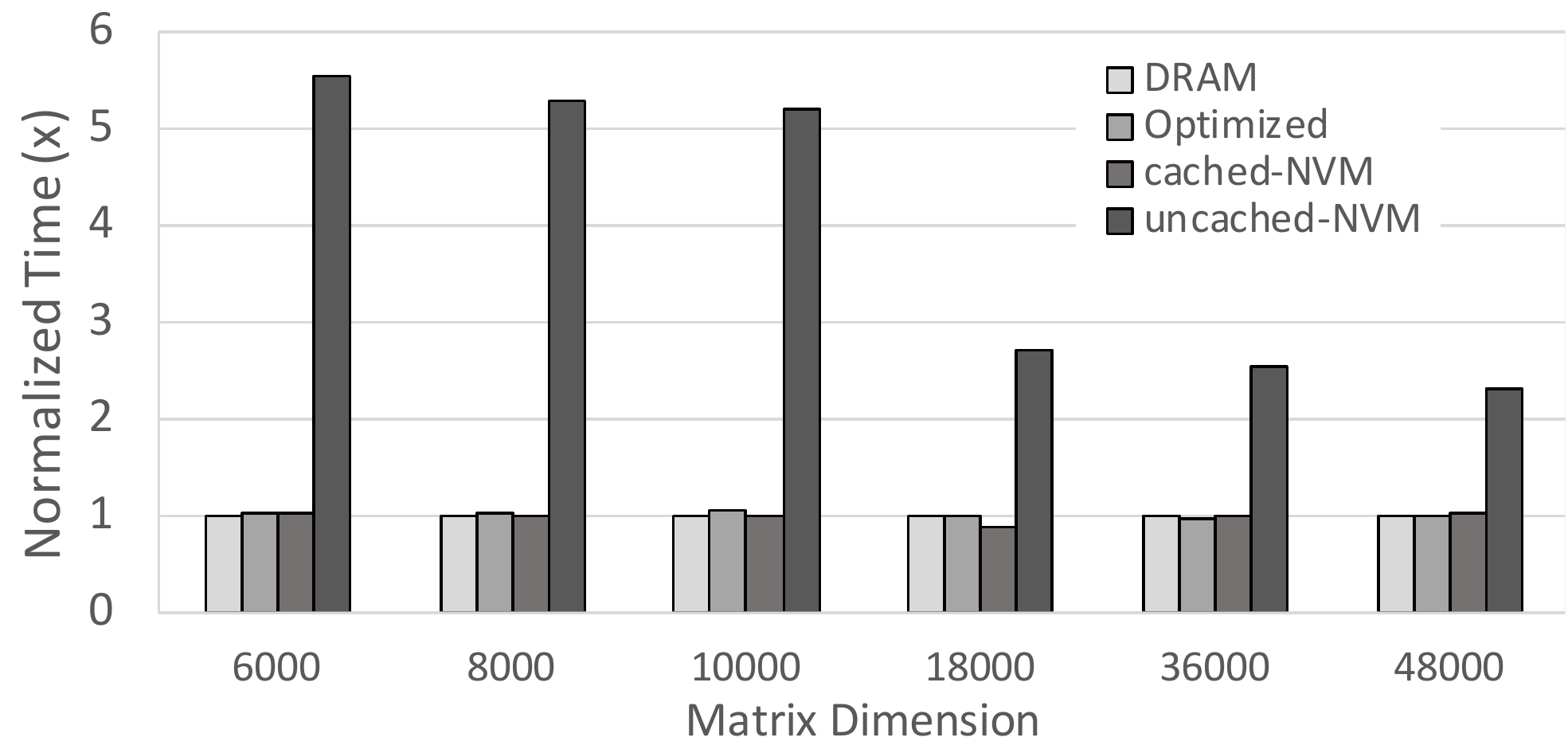}
	\caption{Compare the optimized Scalapack with the original on DRAM, cached-NVM and uncached-NVM.\label{fig:opt}}
\end{figure}

We use a hardware sampling-based implementation of the data-centric profiling tool~\cite{peng2017rthms} to identify write-intensive data structures. Then, the source code is modified accordingly to place the data structures onto DRAM using APIs in~\cite{peng2019}. Figure~\ref{fig:opt} presents the result after applying this optimization in ScaLAPACK, as compared to that on DRAM-only, cached-NVM, and uncached-NVM. The results show that this approach manages to achieve DRAM-like performance at different problem sizes. Note that the used DRAM size in this optimization is only $30\%$ of the DRAM and cached-NVM modes. As a validation, we also placed other read-intensive data structures onto DRAM, which results in little difference compared to that on uncached-NVM.

\section{Related Work}
Before the hardware of byte-addressable NVM becomes available, most previous works use emulators and simulators for evaluating their approaches on NVM~\cite{Mnemosyne:ASPLOS11,nvtree:fast15,Kolli:ASPLOS2016, Pelley:isca14,unimem:sc17,peng2018characterizing,sc18:wu,peng2018siena,Fernando:hpdc18}. Although simulations and emulations can provide valuable insights into the performance trend, they either lack the performance details or are constraint by small problems due to the long simulation time. In~\cite{ucsd19} and~\cite{Yang2019AnEG}, the authors prove that using software emulation or hardware emulation does not capture all the features of real hardware like the Intel Optane. Therefore, the system software for NVM proposed in the previous studies requires re-evaluation. Different from these works, the findings and insights in this work are derived from representative HPC applications on real NVM hardware.

Since the release of real hardware, several works have provided preliminary evaluations of the Intel Optane DC PMM~\cite{ucsd19,peng2019}. Some works have also ported selected applications in commercial database, scientific and graph workloads onto Optane~\cite{utexas19,vanRenen:2019:PMI:3329785.3329930,Yang2019AnEG, Patil:memsys19}. For instance,~\cite{utexas19} optimizes the graph workload Galois to mitigate the NUMA effect when Optane is in the Memory mode. Still, a comprehensive evaluation that covers the landscape of HPC applications as in our work is missing.

Prior works have proposed various approaches for utilizing NVM-based heterogeneous memory systems~\cite{unimem:sc17,sc18:wu,peng2018siena,Fernando:hpdc18}. Unimem~\cite{unimem:sc17} uses a sample-based approach to collect memory access information to decide data placement on NVM-DRAM systems. Siena~\cite{peng2018siena} explores different organizations and configurations of DRAM and NVM in a memory system to decide optimal system designs for HPC applications. NVStream~\cite{Fernando:hpdc18} utilizes the byte-addressability in NVM to remove expensive system calls and uses non-temporary storage and delta compression to reduce overhead due to ensuring crash consistency on NVM. In this work, we identify new optimization priorities and insights that will also benefit these approaches and techniques.  
\section{Conclusion}
In this work, we analyze the performance of the Seven Dwarfs on NVM-based heterogeneous main memory. Our results quantify the effectiveness of using DRAM as a cache to NVM to enable large problems at reasonable performance. For uncached NVM, we identify that the write throttling effect and concurrency contention requires a high priority in optimizations. We highlight that changing concurrency may have a diverging effect on read and write access in some applications. Therefore, a global adjustment of the concurrency may be insufficient. For the cached-NVM, we develop a prediction model based on hardware events collected from a small set of application runs to predict the performance at various concurrency and data size. For uncached-NVM, we demonstrate the effectiveness of write-aware data placement. Overall, our study provides insights and feedback for designing and exploiting NVM-based main memory on future supercomputers.

\section*{Acknowledgment}
\scriptsize{This work was performed under the auspices of the U.S. Department of Energy by Lawrence Livermore National Laboratory under contract No. DE-AC52-07NA27344. This research was supported by the Exascale Computing Project (17-SC-20-SC), a collaborative effort of the U.S. Department of Energy Office of Science and the National Nuclear Security Administration. LLNL-PROC-793886. This work was partially supported by U.S.National Science Foundation (CNS-1617967, CCF-1553645 and CCF-1718194).}



\bibliographystyle{plain}
\bibliography{./main,./li,./kai}

\begin{thebibliography}{10}

\bibitem{asanovic2006landscape}
Krste Asanovic, Ras Bodik, Bryan~Christopher Catanzaro, Joseph~James Gebis,
  Parry Husbands, Kurt Keutzer, David~A Patterson, William~Lester Plishker,
  John Shalf, Samuel~Webb Williams, et~al.
\newblock The landscape of parallel computing research: A view from berkeley.
\newblock Technical report, Technical Report UCB/EECS-2006-183, EECS
  Department, University of California at Berkeley, 2006.

\bibitem{bailey1991parallel}
David~H Bailey, Eric Barszcz, John~T Barton, David~S Browning, Robert~L Carter,
  Leonardo Dagum, Rod~A Fatoohi, Paul~O Frederickson, Thomas~A Lasinski, Rob~S
  Schreiber, et~al.
\newblock {The NAS parallel benchmarks}.
\newblock {\em The International Journal of Supercomputing Applications},
  5(3):63--73, 1991.

\bibitem{bell2012boxlib}
J~Bell, A~Almgren, V~Beckner, M~Day, M~Lijewski, A~Nonaka, and W~Zhang.
\newblock Boxlib user’s guide.
\newblock {\em github. com/BoxLib-Codes/BoxLib}, 2012.

\bibitem{blackford1997scalapack}
L~Susan Blackford, Jaeyoung Choi, Andy Cleary, Eduardo D'Azevedo, James Demmel,
  Inderjit Dhillon, Jack Dongarra, Sven Hammarling, Greg Henry, Antoine
  Petitet, et~al.
\newblock {\em {ScaLAPACK} users' guide}, volume~4.
\newblock Siam, 1997.

\bibitem{curtis2008prediction}
Matthew Curtis-Maury, Filip Blagojevic, Christos~D Antonopoulos, and
  Dimitrios~S Nikolopoulos.
\newblock Prediction-based power-performance adaptation of multithreaded
  scientific codes.
\newblock {\em IEEE Transactions on Parallel and Distributed Systems},
  19(10):1396--1410, 2008.

\bibitem{davis2011university}
Timothy~A Davis and Yifan Hu.
\newblock The university of florida sparse matrix collection.
\newblock {\em ACM Transactions on Mathematical Software (TOMS)}, 38(1):1,
  2011.

\bibitem{dobrev2012high}
Veselin~A Dobrev, Tzanio~V Kolev, and Robert~N Rieben.
\newblock High-order curvilinear finite element methods for lagrangian
  hydrodynamics.
\newblock {\em SIAM Journal on Scientific Computing}, 34(5):B606--B641, 2012.

\bibitem{Fernando:hpdc18}
Pradeep Fernando, Ada Gavrilovska, Sudarsun Kannan, and Greg Eisenhauer.
\newblock Nvstream: Accelerating hpc workflows with nvram-based transport for
  streaming objects.
\newblock In {\em Proceedings of the 27th International Symposium on
  High-Performance Parallel and Distributed Computing}, HPDC '18, 2018.

\bibitem{utexas19}
Gurbinder Gill, Roshan Dathathri, Loc Hoang, Ramesh Peri, and Keshav Pingali.
\newblock Single machine graph analytics on massive datasets using intel optane
  dc persistent memory, 2019.

\bibitem{habib2013hacc}
Salman Habib, Vitali Morozov, Nicholas Frontiere, Hal Finkel, Adrian Pope, and
  Katrin Heitmann.
\newblock {HACC}: extreme scaling and performance across diverse architectures.
\newblock In {\em Proceedings of the International Conference on High
  Performance Computing, Networking, Storage and Analysis}, page~6. ACM, 2013.

\bibitem{hosomi2005novel}
M~Hosomi, H~Yamagishi, T~Yamamoto, K~Bessho, Y~Higo, K~Yamane, H~Yamada,
  M~Shoji, H~Hachino, C~Fukumoto, et~al.
\newblock A novel nonvolatile memory with spin torque transfer magnetization
  switching: Spin-ram.
\newblock In {\em IEEE InternationalElectron Devices Meeting, 2005. IEDM
  Technical Digest.}, pages 459--462. IEEE, 2005.

\bibitem{ucsd19}
Joseph Izraelevitz, Jian Yang, Lu~Zhang, Juno Kim, Xiao Liu, Amirsaman
  Memaripour, Yun~Joon Soh, Zixuan Wang, Yi~Xu, Subramanya~R. Dulloor, Jishen
  Zhao, and Steven Swanson.
\newblock Basic performance measurements of the intel optane dc persistent
  memory module, 2019.

\bibitem{Kolli:ASPLOS2016}
Aasheesh Kolli, Steven Pelley, Ali Saidi, Peter~M. Chen, and Thomas~F. Wenisch.
\newblock {High-Performance Transactions for Persistent Memories}.
\newblock In {\em Proceedings of the Twenty-First International Conference on
  Architectural Support for Programming Languages and Operating Systems}, 2016.

\bibitem{aurora_anl}
Argonne~National Lab.
\newblock {{U.S. Department of Energy and Intel to deliver first exascale
  supercomputer}}.
\newblock
  https://www.anl.gov/article/us-department-of-energy-and-intel-to-deliver-first-exascale-supercomputer.

\bibitem{lee2009architecting}
Benjamin~C Lee, Engin Ipek, Onur Mutlu, and Doug Burger.
\newblock Architecting phase change memory as a scalable dram alternative.
\newblock {\em ACM SIGARCH Computer Architecture News}, 37(3):2--13, 2009.

\bibitem{nvm_ipdps12}
Dong Li, Jeffrey Vetter, Gabriel Marin, Collin McCurdy, Cristi Cira, Zhuo Liu,
  and Weikuan Yu.
\newblock {Identifying Opportunities for Byte-Addressable Non-Volatile Memory
  in Extreme-Scale Scientific Applications}.
\newblock In {\em International Parallel and Distributed Processing Symposium},
  2012.

\bibitem{li2005overview}
Xiaoye~S Li.
\newblock An overview of superlu: Algorithms, implementation, and user
  interface.
\newblock {\em ACM Transactions on Mathematical Software (TOMS)},
  31(3):302--325, 2005.

\bibitem{Patil:memsys19}
Onkar Patil, Latchesar Ionkov, Jason~Lee Lee, Frank Mueller, and Michael~Lang
  Lang.
\newblock Performance characterization of a dram-nvm hybrid memory architecture
  for hpc applications using intel optane dc persistent memory modules.
\newblock In {\em Proceedings of the International Symposium on Memory
  Systems}, MEMSYS '19, 2019.

\bibitem{Pelley:isca14}
Steven Pelley, Peter~M. Chen, and Thomas~F. Wenisch.
\newblock {Memory Persistency}.
\newblock In {\em ISCA}, 2014.

\bibitem{peng2018siena}
I.~B. {Peng} and J.~S. {Vetter}.
\newblock Siena: Exploring the design space of heterogeneous memory systems.
\newblock In {\em SC18: International Conference for High Performance
  Computing, Networking, Storage and Analysis}, pages 427--440, Nov 2018.

\bibitem{peng2019}
Ivy~B. Peng, Maya~B. Gokhale, and Eric~W. Green.
\newblock System evaluation of the intel optane byte-addressable {NVM}.
\newblock In {\em Proceedings of the International Symposium on Memory
  Systems}. ACM, 2019.

\bibitem{peng2017rthms}
Ivy~Bo Peng, Roberto Gioiosa, Gokcen Kestor, Pietro Cicotti, Erwin Laure, and
  Stefano Markidis.
\newblock Rthms: A tool for data placement on hybrid memory system.
\newblock In {\em ACM SIGPLAN Notices}, volume~52, pages 82--91. ACM, 2017.

\bibitem{peng2018characterizing}
Ivy~Bo Peng, Roberto Gioiosa, Gokcen Kestor, Jeffrey~S Vetter, Pietro Cicotti,
  Erwin Laure, and Stefano Markidis.
\newblock Characterizing the performance benefit of hybrid memory system for
  hpc applications.
\newblock {\em Parallel Computing}, 76:57--69, 2018.

\bibitem{Qureshi2009}
Moinuddin~K. Qureshi, Vijayalakshmi Srinivasan, and Jude~A. Rivers.
\newblock Scalable high performance main memory system using phase-change
  memory technology.
\newblock In {\em Proceedings of the 36th Annual International Symposium on
  Computer Architecture}, ISCA '09, pages 24--33, New York, NY, USA, 2009. ACM.

\bibitem{strukov2008missing}
Dmitri~B Strukov, Gregory~S Snider, Duncan~R Stewart, and R~Stanley Williams.
\newblock The missing memristor found.
\newblock {\em nature}, 453(7191):80, 2008.

\bibitem{pcm}
{Thomas Willhalm, Roman Dementiev, Patrick Fay}.
\newblock Intel performance counter monitor - a better way to measure cpu
  utilization, 2017.

\bibitem{tramm2014xsbench}
John~R Tramm, Andrew~R Siegel, Tanzima Islam, and Martin Schulz.
\newblock Xsbench-the development and verification of a performance abstraction
  for monte carlo reactor analysis.
\newblock {\em The Role of Reactor Physics toward a Sustainable Future
  (PHYSOR)}, 2014.

\bibitem{vanRenen:2019:PMI:3329785.3329930}
Alexander van Renen, Lukas Vogel, Viktor Leis, Thomas Neumann, and Alfons
  Kemper.
\newblock Persistent memory {I/O} primitives.
\newblock In {\em Proceedings of the 15th International Workshop on Data
  Management on New Hardware}, DaMoN'19, 2019.

\bibitem{Mnemosyne:ASPLOS11}
Haris Volos, Andres~Jaan Tack, and Michael~M. Swift.
\newblock Mnemosyne: Lightweight persistent memory.
\newblock In {\em Proceedings of the Sixteenth International Conference on
  Architectural Support for Programming Languages and Operating Systems},
  ASPLOS, 2011.

\bibitem{unimem:sc17}
Kai Wu, Yingchao Huang, and Dong Li.
\newblock {Unimem: Runtime Data Management on Non-volatile Memory-based
  Heterogeneous Main Memory}.
\newblock In {\em SC}, 2017.

\bibitem{sc18:wu}
Kai Wu, Jie Ren, and Dong Li.
\newblock Runtime data management on non-volatile memory-based heterogeneous
  memory for task-parallel programs.
\newblock In {\em Proceedings of the International Conference for High
  Performance Computing, Networking, Storage, and Analysis}, 2018.

\bibitem{Yang2019AnEG}
Jianhua Yang, Juno Kim, Morteza Hoseinzadeh, Joseph Izraelevitz, and Steven
  Swanson.
\newblock An empirical guide to the behavior and use of scalable persistent
  memory.
\newblock {\em ArXiv}, abs/1908.03583, 2019.

\bibitem{nvtree:fast15}
Jun Yang, Qingsong Wei, Cheng Chen, Chundong Wang, Khai~Leong Yong, and
  Bingsheng He.
\newblock {NV-Tree}: Reducing consistency cost for nvm-based single level
  systems.
\newblock In {\em 13th {USENIX} Conference on File and Storage Technologies
  ({FAST} 15)}, 2015.

\bibitem{yang2006parallel}
Ulrike~Meier Yang.
\newblock Parallel algebraic multigrid methods—high performance
  preconditioners.
\newblock In {\em Numerical solution of partial differential equations on
  parallel computers}, pages 209--236. Springer, 2006.

\end{thebibliography}
%

\end{document}